\documentclass[12pt]{article}
\usepackage{graphicx}
\begin{document}

\title{$CP$-violation in $K_{S,L}\rightarrow \pi^{+}\pi^{-}\gamma$
and $K_{S,L}\rightarrow \pi^{+}\pi^{-}e^{+}e^{-}$ decays}
\author{S. S. Bulanov\\
{\small Institute of Theoretical and Experimental Physics,}\\
{\small B. Cheremyshkinskaya ul., 25, 117259 Moscow, Russia.}\\
{\small e-mail: bulanov@heron.itep.ru}
}
\date{}
\maketitle

\begin{abstract}
The dependence of $K_{S,L}\rightarrow \pi^{+}\pi^{-}\gamma$
decay probabilities on photon polarization is calculated.
The phases of terms of amplitude that arise from the pion-pion interaction are
obtained by using a simple realistic model of pion-pion interaction via
virtual $\rho$-meson, instead of the ChPT. The results are compared
with those of other authors and the origin of the discrepancies is explained.
It is shown that the standard ChPT approach for $K_{S,L}\rightarrow 
\pi^{+}\pi^{-}\gamma$ decays cannot reproduce the contribution
of the $\rho$-meson to the $P$-wave $\pi\pi$ interaction. The departure of the 
photon spectrum from pure bremsstrahlung due to the pion loop contribution to 
the direct emission amplitude is calculated. It is shown that the interference
between the terms of amplitude with different $CP$-parity appears only when
the photon is polarized (linearly or circularly). Instead of measuring the
linear polarization, the angular correlation between the $\pi^{+}\pi^{-}$ and
$e^{+}e^{-}$ planes in $K_{S,L}\rightarrow\pi^{+}\pi^{-}e^{+}e^{-}$ decay can
be studied.

\end{abstract}

\section{Introduction}

The theoretical and experimental study of the $CP$-violation in
the radiative decays of the $K_L$ and $K_S$ has a long history. In
view of future precise measurements of these decays we have
recalculated the above effects. Generally our results are in
agreement with the previous ones. A few discrepancies (see Conclusion)
are caused by more realistic evaluation of pion loops in the
present paper.

The pattern of the CP-violation in the $K_{L,S}\rightarrow \pi ^{+}\pi
^{-}\gamma$ decays was theoretically predicted in the 1960s. H. Chew \cite{bib
Chew} determined the amplitude structure of $K_{1,2}^0\rightarrow \pi ^{+}\pi
^{-}\gamma$ decays. He calculated the pion loop contribution to the direct
emission amplitude and stated the possibility of the $CP$-violation in case the
amplitude is a sum of terms with different $CP$-parity. G. Costa and P. K.
Kabir \cite{bib 2a} as well as L. M. Sehgal and L. Wolfenstein \cite{SW}
studied the interference of the $K_1^0$ and $K_2^0$ in the decays into $\pi
^{+}\pi ^{-}\gamma$, identifying this effect with the $CP$-violation. They
also qualitatively discussed the dependence of the decay probability on the
photon polarization. A. D. Dolgov and L. A. Ponomarev \cite{bib dolgov} paid
special attention to the $K_L$ decay. They realized that the $CP$-violation
effects in the $K_L$ decay should be larger than in the $K_S$ decay. They also
found that the measurement of the photon polarization could enhance the
signals of the $CP$-violation. They qualitatively discussed the measurement of
the angular correlation between the $\pi^{+}\pi^{-}$ and  $e^{+}e^{-}$
planes in $K_L\rightarrow\pi^{+}\pi^{-}e^{+}e^{-}$
decay instead of measuring the linear polarization.

In the 1990s $K_{L,S}\rightarrow \pi ^{+}\pi ^{-}\gamma
$ decays were thoroughly studied using ChPT (Chiral Perturbation Theory)
with special emphasize on the $K_L$-meson decay. The $K_L$ decay attracted
special attention, because, as it is known experimentally, the contributions
of the $K_L$ decay
amplitude terms with different $CP$-parity are of comparable
magnitude and this makes the $CP$-violation to be distinctively seen.
Contrary to this in the case of the $K_S$ decay the $CP$-violation is
difficult to detect due to the fact that the internal bremsstrahlung
contribution shades the contribution of the direct emission.

G. D'Ambrosio and G. Isidori \cite{AmbrI} and  G. D'Ambrosio, M.
Miragluiolo and P. Santorelli \cite{bib 1} presented a
complete calculation of the direct emission contribution to the
$K_{S,L}\rightarrow \pi ^{+}\pi ^{-}\gamma$ decay amplitude up to
the 6th order of momenta in the framework of the ChPT. In view of
future precise measurements and the new data on direct
$CP$-violation obtained by KTeV collaboration (A. Alavi-Harati {\it et. al.})
\cite{CP in Kpp} several authors addressed the problem of short
distance contributions to the direct $CP$-violating observables in
the radiative $K$-meson decays. X.-G. He and G. Valencia \cite{He}
studied the $s\rightarrow d\gamma$ transition. They described the
long distance contribution in the framework of the ChPT and subtracted it
from the physical amplitudes of the $K\rightarrow\pi\pi\gamma$
decays in order to constrain the new, short distance,
interactions. They also illustrated two types of models in which the short
distance
interactions could be significantly enhanced with respect to the
Standard Model, namely the left-right symmetric model and the
supersymmetry.

G. Colangelo, G. Isidori and J. Portoles \cite{Colangelo}
analyzed the supersymmetric contributions to the direct
$CP$-violating observables in $K\rightarrow\pi\pi\gamma$ decays
induced by gluino-mediated magnetic-penguin operators. They found
that the direct $CP$-violation could be substantially enhanced
with respect to its Standard Model value especially in the
scenario where the direct $CP$-violation is dominated by
supersymmetric contributions.

J. Tandean and G. Valencia \cite{bib valencia}
also revisited the $K_L\rightarrow \pi ^{+}\pi ^{-}\gamma$ decay
in order to study the possible contributions of $s\rightarrow d\gamma$
as well as gluonic, $s\rightarrow dg$, transitions to the direct
$CP$-violating observables in the framework of two models: left-right
symmetric model and the supersymmetry.

The experimental studies of the $K_L$ decay were reported in Refs.
\cite{bib carrol}, \cite{bib ramberg}, \cite{bib kettell}. In 1980
A. S. Carrol {\it et. al}. \cite{bib carrol} first observed both the internal
bremsstrahlung
and the direct emission contributions in the $K_L$ decay. E. J. Ramberg {\it
et. al}
\cite{bib ramberg} in 1993 presented more precise measurements of the pattern
of the
$CP$-violation in the $K_L$ decay. S. Kettell  \cite{bib kettell}
in his talk "Experimental Results on Radiative Kaon Decays"
summarized the recent more precise results.

In the case of the $K_S$ decay only the internal bremsstrahlung
contribution was found by E. J. Ramberg {\it et. al}  \cite{bib ramberg}
and  by H. Taureg {\it et. al}  \cite{taureg}. The latter established the
upper bounds
for the branching ratio of the interference of internal bremsstrahlung
and electric direct emission.
The theoretical study of the $K_S$ decay was performed by
G. D'Ambrosio, M. Miragliuolo and F. Sannino \cite{bib 3},
where the electric direct emission amplitude was calculated in the
framework of the ChPT. They also studied the photon spectrum
departure from the pure internal bremsstrahlung expectation due to
the interference of the internal bremsstrahlung and the electric direct
emission. In Ref. \cite{bib 3} there was also noted that the
measurement of this interference would provide a test of the
proposed models.

In regard to the future higher precision experiments on the $CP$-violation
it is important to study in more detail the properties of the phases
caused by the $\pi\pi$-interaction in the $K_{L,S}\rightarrow \pi ^{+}\pi
^{-}\gamma$ decays.
More precise calculation of these phases is the prerequisite for extracting
the precise values of the $CP$-violating parameters in the
$K$-meson decays.

In the present paper we calculate the probability of the
$K_{S,L}\rightarrow \pi^{+}\pi^{-}\gamma$ decays, using instead of ChPT
a simple realistic model of $\pi\pi$ interaction via virtual $\rho$-meson,
proposed
by B. W. Lee and M. T. Vaughn \cite{bib Lee} for the purposes of studying the
$P$-wave
resonance in the $\pi\pi$ scattering. This model
was elaborated in Ref. \cite{bib Chew} for the purposes of
describing the electric direct emission in the $K\rightarrow 2\pi\gamma$
decays. In the framework of this model we shall derive the
phases of the amplitude terms, connected with the $\pi\pi$-interaction.
According to the approach used in Ref. \cite{bib Chew}
we shall show that the phase of the electric
direct emission amplitude is not equal to the phase of the $\pi\pi$
scattering in the $P$-wave, which could be expected according to the
final state interaction theorem formulated by K. Watson \cite{bib fsi}, E.
Fermi
\cite{Fermi}, S. Fubini and Y. Nambu \cite{Fubini}.
This happens due to the fact that the interaction of
pions occurs not in the final state, but in the intermediate state.

We shall calculate the departure of the photon spectrum from bremsstrah-lung
due to the pion loop contribution to the direct emission amplitude in
the $K_S$ decay. Then we shall compare our results on the photon energy
dependence of the interference between the internal bremsstrahlung and the
electric direct emission in the $K_S$ decay with the results of Refs.
\cite{AmbrI}, \cite{bib 3} obtained in the framework of ChPT. It will be
proved that the "interference branching ratio" differs from the
the one obtained in Refs. \cite{AmbrI}, \cite{bib 3}
(see Table 1) and that the standard ChPT approach even with
higher order counterterms taken into account cannot reproduce the
contribution of the $\rho$-meson. If the counterterm
contributions are set to zero ($k_f=0$), the
result of the present paper and the one obtained in  Refs. \cite{AmbrI},
\cite{bib 3} for the "interference branching ratio" are in agreement for the
photon energy cut $\omega>20$ MeV. However, for the photon energy cuts
$\omega>50$ MeV and $\omega>100$ MeV the discrepancy appears. It is due to the
fact that the photon spectra of the interference contributions differ. This can
be clearly seen from Fig. 5, where the result of the present paper (solid
curve) along with the results of Refs. \cite{AmbrI}, \cite{bib 3} (dashed
curves, $k_f=-0.5,~0,+0.5$) for the photon spectra of the interference
contributions are shown.

If the counterterm contributions are switched on, then we shall see that the
arising discrepancy is rather large for given in Refs. \cite{AmbrI}, \cite{bib
3} values of the counterterm contributions (e.g. for $k_f=0.5$ the result of
the present paper is 1.7 times smaller than that of Refs. \cite{AmbrI},
\cite{bib 3}, while for $k_f=-0.5$ it is 3.3 times larger, see Table 1). It
should be stressed that the counterterm contributions do not depend on the
photon energy and exhibit the behavior different from the results of the
present paper (see Fig. 5).We shall show that the discrepancy
between these results appears due to the difference in models used
to calculate the amplitude of $\pi\pi$ scattering in $P$-wave.
We shall notice that this difference appears due to the fact that
in Refs. \cite{AmbrI}, \cite{bib 3} the $\rho$-meson contribution
shows up only in low-energy constants, while the resonance
behavior was not reproduced. Such an approach did not take
into account the photon energy dependence of the $\delta_1^1$
phase, i. e. the behavior of the $\rho$ propagator. Instead, we
shall take into account the additional phase shift produced by pion loops
and the energy dependence of the $P$-wave $\pi\pi$ scattering phase.
We shall use the approach of Refs. \cite{bib Chew}, \cite{bib Lee}
in order to describe the phase energy dependence, which
appears in the amplitude through the pion loop contributions to the
$\rho$-meson propagator.

In order to show that the inclusion of the $\rho$-meson into the
analysis of the electric direct emission in the $K\rightarrow\pi\pi\gamma$
decays is important, we shall address the problem of $\pi\pi$
scattering in the $P$-wave.
We shall compare the experimental data with the results obtained
in the framework of different models: a) the ChPT, b) the ChPT with 
$\rho$-meson contribution taken into account and c) the simple realistic model.
We shall see that the results calculated within the framework of the ChPT
coincide with the experimental data only for low energies, because
the ChPT doesn't take into account the resonance contribution of
the $\rho$-meson. Then we shall see that the model of
$\pi\pi$-interaction via $\rho$-meson and the ChPT with
$\rho$-meson contribution show the same
behavior of the $P$-wave $\pi\pi$ scattering phase.

In regard to the effects of $CP$-violation in the
$K_{S,L}\rightarrow\pi^{+}\pi^{-}\gamma$ decays we shall note that
in the case of the $K_S$ decay the contribution of the
$CP$-violating magnetic direct emission to the decay probability
is negligibly small contrary to the case of the $K_L$ decay, where the
contributions of the $CP$-conserving  magnetic direct emission and
$CP$-violating internal bremsstrahlung to the decay probability
are of comparable magnitude as it was theoretically predicted in
Refs. \cite{AmbrI}, \cite{bib 1}, \cite{bib valencia},
and experimentally confirmed in Refs. \cite{bib carrol},
\cite{bib ramberg}, \cite{bib kettell}. Therefore, in the case of
the $K_S$ decay we shall consider the interference between
the amplitude terms with different $CP$-parity, due to the fact that
the interference is the largest term in which the $CP$-violating
effects reside. As we shall see one has to measure the polarization of photons
to analyze the interference, because it is nonzero only when the
polarization of the photon is observed. We note that this
phenomenon was discussed qualitatively by G.
Costa and P. K. Kabir \cite{bib 2a}, L. M. Sehgal and L. Wolfenstein \cite{SW},
\cite{bib dolgov}, A. D. Dolgov and L. A. Ponomarev \cite{bib dolgov},
M. McGuigan and A. I. Sanda \cite{bib sanda} for both $K_S$ and $K_L$
decays. The quantitative analysis of the $K_L$ decay amplitude dependence
on the photon polarization was performed by L. M. Sehgal and J. van Leusen
\cite{bib Sehgal1},
\cite{sehgal2}, \cite{sehgal3}.

According to the approach proposed in Ref. \cite{bib Sehgal1} for $K_L$ decay,
we shall examine the $K_S\rightarrow \pi^{+}\pi^{-}\gamma$ decay
probability with the polarized photon, taking into account various cases of
the photon
polarization and we shall show that the measurement of the linear polarization
in principle allows extraction of terms with opposite $CP$-parity.
In the present paper we address the problem of studying the effects of
different cases of the photon polarization in the case of the $K_S$
decay. As it is known as an alternative to measuring the linear photon
polarization,
the angular correlation of $\pi^{+}\pi^{-}$ and $e^{+}e^{-}$
planes in $K_{S,L}\rightarrow\pi^{+}\pi^{-}e^{+}e^{-}$
decay can be studied, as it was first suggested in Ref. \cite{bib
dolgov}. The structure of the $K_L\rightarrow\pi^{+}\pi^{-}e^{+}e^{-}$ decay
amplitude was studied by L. M. Sehgal, M. Wanninger \cite{bib sehgal} P.
Heiliger,
L. M. Sehgal \cite{heiliger} and J. K. Elwood, M. B. Wise \cite{elwood}, where
the
$CP$-violating asymmetry, arising from the angular correlation
of $\pi^{+}\pi^{-}$ and $e^{+}e^{-}$ planes, was also obtained. The
predictions of Refs.
\cite{bib sehgal}, \cite{heiliger}, \cite{elwood} on the
$K_L\rightarrow\pi^{+}\pi^{-}e^{+}e^{-}$
decay branching ratio and $CP$-violating asymmetry
were confirmed by KTeV collaboration (A. Alavi-Harati {\it et. al.}) \cite{bib
ktev}.

The problem of pion loop contribution in the
$K_L\rightarrow\pi^{+}\pi^{-}e^{+}e^{-}$
decay was studied by J. K. Elwwod, M. B. Wise, M. J. Savage and J. W. Walden
\cite{Savage1}
in the framework of the ChPT. Their approach included both $\pi\pi\rightarrow
\pi\pi$
and $\pi\pi\rightarrow \pi\pi\gamma^{*}$ rescattering. Previous estimates of
the effect of
pion loop contribution in Refs. \cite{bib sehgal}, \cite{heiliger} used the
measured
pion phase shifts and neglected $\pi\pi\rightarrow \pi\pi\gamma^{*}$.
They found that this effect could enhance the $CP$-violating
asymmetry by about 45\% over the estimates given in Ref. \cite{elwood},
resulting
in the value of the asymmetry, which is in good agreement with
the experimental data \cite{bib ktev}. These results were summarized by M. J.
Savage in his
talk "$K_L^0\rightarrow\pi^{+}\pi^{-}e^{+}e^{-}$ in Chiral Perturbation
Theory" \cite{Savage2}.
A more precise calculation of the $K_L\rightarrow\pi^{+}\pi^{-}e^{+}e^{-}$
decay branching
ratio up to next-to-leading order in the framework of the
ChPT was presented by H. Pichl \cite{Pichl1}. G. Ecker and H. Pichl
\cite{Pichl2}
updated the theoretical analysis of the $CP$-violating asymmetry in the
$K_L\rightarrow\pi^{+}\pi^{-}e^{+}e^{-}$ decay using the ChPT and the most
recent
phenomenological information.

In the present paper we shall calculate the $CP$-violating asymmetry in
the case of the $K_L\rightarrow\pi^{+}\pi^{-}e^{+}e^{-}$ decay in order to
compare our results with
those of Refs. \cite{Savage1}, \cite{Savage2}. The result of Refs.
\cite{Savage1}, \cite{Savage2}
for the $CP$-violating asymmetry is 14\%, and we shall find that the asymmetry
is $(13.4\pm 0.9)\%$.
The central values coincide within the accuracy of the calculation.
We shall also calculate the $CP$-violating asymmetry in
the case of the $K_S\rightarrow\pi^{+}\pi^{-}e^{+}e^{-}$ decay and we shall
find it
to be substantially smaller: $(5.1\pm 0.4)\times 10^{-5}$.
This could be qualitatively expected from the analysis
of the $K_S\rightarrow \pi^{+}\pi^{-}\gamma$ decay, because the $CP$-violating
asymmetry
depends on the interference of the amplitude terms with opposite
$CP$-parity. The $CP$-violating magnetic direct emission amplitude
is very small in the $K_S$ decay contrary to that of the $K_L$,
where the $CP$-conserving magnetic emission and $CP$-violating
internal bremsstrahlung are of comparable magnitude.
In order to detect the asymmetry experimentally one will need more than
$10^{10}$ $K_S\rightarrow\pi^{+}\pi^{-}e^{+}e^{-}$ decays, because
one should have the statistical error to be smaller than the magnitude of the
effect.

The paper is organized as follows. In Section 2 we discuss the
structure of the $K_{S,L}\rightarrow \pi^{+}\pi^{-}\gamma$ decay
amplitude. In Section 3 we present the simple realistic model of $\pi\pi$
scattering in the $P$-wave and derive the expression for pion loops
contribution to the direct emission amplitude in the framework of this model.
We discuss the
spectrum departure from pure internal bremsstrahlung due to pion
loops contribution in $K_S$ decay in Section 4. We compare our
results on the spectrum departure with the results obtained in the
framework of ChPT in Section 5. In Section 6 we discuss the
different models of the $\pi\pi$ scattering in $P$-wave and compare
them with experimental data. In Section 7 we carry out the
analysis of the dependence of the $K_{S,L}$ decay amplitude on the
photon polarization in terms of Stokes vectors. We also discuss
the angular correlation in $K_{S,L}\rightarrow\pi^{+}\pi^{-}e^{+}e^{-}$
decays and calculate the $CP$-violating asymmetry in
$K_S\rightarrow\pi^{+}\pi^{-}e^{+}e^{-}$ decay in this section.
Section 8 is devoted to the discussion of the main results and conclusion.

\section{The amplitude structure}

We start by labeling the momenta of the particles involved in the decay
\begin{equation}
K_{S,L}(r)\rightarrow\pi^{+}(p)\pi^{-}(q)\gamma(k,e),
\end{equation}
where $e$ is a 4-vector of photons wave function.

It is convenient to define three expressions:
\begin{equation}
T_{B}=\frac{pe}{pk}-\frac{qe}{qk}, \label{eq.IB}
\end{equation}
\begin{equation}
T_{E}=(pe)(qk)-(qe)(pk), \label{eq.E}
\end{equation}
\begin{equation}
T_{M}=\varepsilon_{\mu\nu\rho\sigma}p_\mu q_\nu k_\rho e_\sigma.
\label{eq.M}
\end{equation}

The amplitudes of $K_{S,L}\rightarrow \pi^{+}\pi^{-}\gamma$
decays are made up of two components: the internal bremsstrahlung ($B$),
proportional to $T_B$, and direct emission ($D$) \cite{bib 2a}, \cite{SW},
\cite{bib 1}. In its
turn, $D$ is a sum of an electric term($E_{D}$), proportional to
$T_E$, and a magnetic term($M_{D}$), proportional to $T_M$.
We note that  $T_E=T_B (pk)(qk)$; however, due to the different origin
of the internal bremsstrahlung and electric direct emission it is
convenient to treat them separately.

In accordance with the above, the amplitudes of the
$K_{S,L}\rightarrow \pi^{+}\pi^{-}\gamma$ decays can be written as
follows
\begin{equation}
A(K_S\rightarrow \pi^{+}\pi^{-}\gamma)=
eAe^{i\delta_{0}^0}T_{B}+e(a+b)T_E
+ie\eta_{+-}c T_M,   \label{eq.2}
\end{equation}
\begin{equation}
A(K_L\rightarrow \pi^{+}\pi^{-}\gamma)=
\eta_{+-}eA e^{i\delta_{0}^0}T_{B}
+e\eta_{+-}(a+b)T_E
+i~ecT_M,   \label{eq.3}
\end{equation}
where $\delta_{0}^0$ is the $S$-wave pion scattering phase. Here
the upper index is isospin, the lower -- angular momentum and
$\eta_{+-}$ is the well known CP-violation parameter
in the $K_L\rightarrow\pi^{+}\pi^{-}$ decay.
The imaginary unit in front of the factor $c$ stems from the hermiticity of the
Hamiltonian describing direct emission, neglecting final state
interaction. The factor $A\equiv A(K\rightarrow\pi^{+}\pi^{-})$ is
determined by the Low theorem for bremsstrahlung \cite{bib 2}.
The term $(a+b)$ is the electric direct emission coupling, while $c$ is the
magnetic direct emission coupling.

As follows from Eqs. (\ref{eq.2}, \ref{eq.3})
the electric direct emission coupling is divided into two terms.
The first term ($a$) describes the loops of heavy particles. The
second term ($b$) describes the loops of pions. Such subdivision is convenient
because the pion loops contribution has an
absorptive part and hence a phase, contrary to the contribution of the heavy
particle
loops.

The phases of direct emission couplings $a$ and $c$
are dictated by the final state interaction theorem
\cite{bib fsi}, \cite{Fermi}, \cite{Fubini}, \cite{bib 11}.
According to the law of conservation of angular momentum we have
$J_\gamma=J_{\pi\pi}=1$. Here $J_\gamma$ is the total angular momentum of the
photon,
$J_{\pi\pi}$ is the total angular momentum of two pions. Since
$J_{\pi\pi}=l_{\pi\pi}$, where $l_{\pi\pi}$ is the orbital momentum of the two
pions,
the spatial part of the two-pion wave function should be
antisymmetric, the isospin part of the wave function should
also be antisymmetric, according to the Bose generalized
principles; i.e. $J_{\pi\pi}=1$($P$-wave), $T=1$,
where $T$ is isospin of two pions. As a result we have for the
phases
$$
a=|a|e^{i\delta_1^1},~~c=|c|e^{i\delta^1_1},
$$
where $\delta_1^1$ is the pion P-wave scattering phase.

\section{Contribution from pion loops}

The phases of the pion loops and  bremsstrahlung contributions
in equations (\ref{eq.2}), (\ref{eq.3})
are defined by the strong interaction of pions. Let us describe
the simple realistic model of the $\pi\pi$ interaction, mentioned above,
which we shall use while considering the $\pi\pi$ scattering in
the $P$-wave. The Lagrangian of this model \cite{bib Lee} has the
following form
\begin{equation}
L=-\frac{g}{\sqrt{2}}\varepsilon_{ijk}(\phi^i\partial_{\mu}{\phi^j}
-\partial_{\mu}{\phi^i}\phi^j)B^k_{\mu},
\end{equation}
where $i,j,k$ are isotopic indices, $\phi$ is the pion field, $B^k_\mu$ --
$\rho$-meson field.
The $P$-wave resonance in $\pi\pi$ scattering is due to the
resonant structure of the $\rho$-meson propagator, the relevant
part of which is
\begin{eqnarray}
D^{\mu\nu}(k)=-D(k^2)g^{\mu\nu}, \nonumber\\
D(k^2)=[k^2-m_{\rho}^2-\Sigma(k^2)]^{-1},
\end{eqnarray}
where $\Sigma(k^2)$ is the $\rho$ self energy operator; in the "resonance
approximation" in which we consider only the sum of the iterated
bubble diagrams with pions running in the loop, $\Sigma(k^2)$ is
given by the following expression
\begin{equation}
\Sigma(s)=J(s)-J(m_{\rho}^2)+i~Im(\Sigma(s)) \label{sigma},
\end{equation}
with
\begin{equation}
Im
(\Sigma(s))=-\frac{g^2}{48\pi}\frac{(s-4m_\pi)^{3/2}}{s^{1/2}}\theta(s-4m_\pi),
\end{equation}
\begin{equation}
J(s)=\frac{g^2}{48\pi}\left\{\frac{(s-4m_\pi)^{3/2}}{s^{1/2}}
\ln\left[\frac{s^{1/2}+(s-4m_\pi)^{1/2}}{s^{1/2}-(s-4m_\pi)^{1/2}}\right]
-\xi s\right\},
\end{equation}
\begin{equation}
\xi=\frac{m_\rho^2-4m_\pi^2}{m_\rho^2}+\frac{m_\rho^2+2m_\pi^2}{m_\rho}
\frac{(m_\rho^2-4m_\pi^2)^{1/2}}{m_\rho}\ln\left[\frac{m_\rho+
(m_\rho^2-4m_\pi^2)}{m_\rho-(m_\rho^2-4m_\pi^2)}\right].
\end{equation}
Here $\theta(x)$ is the step function $\theta(x)=0$ for $x<0$ and
$\theta(x)=1$ for $x>0$.

In the case of internal bremsstrahlung contribution to the $K_{S,L}$
decay probability the interaction of pions can be
described by the diagrams shown in Fig. 1. Though pions are
in $P$-wave in the final state, as it was shown in the previous
section, this group of diagrams results in the $\delta_0^0$ phase of the
amplitude, as it was assumed in Refs.  \cite{bib 1}, \cite{AmbrI},
\cite{bib valencia}, \cite{bib 3}, \cite{bib sanda}, \cite{bib Sehgal1},
\cite{bib
sehgal}. It is due to the fact that the interaction of pions occurs not in the
final
state, but in the intermediate. The diagrams of Fig. 1 contribute
to the $K\pi\pi$ vertex, these corrections are taken into account
by using the the amplitude of the $K_S\rightarrow\pi^{+}\pi^{-}$ decay
as the interaction constant and assuming it equal to its
experimental value.

The emission of the photon from the loops of pions
is governed by another group of diagrams shown in Fig. 2.
Each of these diagrams is divergent but their sum is
finite. The similar result holds in the ChPT.
A straightforward calculation of this finite expression
gives the result that is not gauge invariant. This effect
arises from the cancellation of two 4-dimensional integrals
proportional to $l^2$ and $l_\mu l_\nu$. The evaluation of these
two integrals in the framework of the dimensional regularization
scheme leads to a constant term that restores the gauge invariance \cite{bib
Roma}.

Further we shall neglect the energy dependence of the
$K\pi\pi$ vertex. We shall take the amplitude of the
$K_S\rightarrow\pi^{+}\pi^{-}$ decay as the interaction constant,
and use the experimental result for it.

We find for the matrix element arising from the diagrams, shown in  Fig. 2:
\begin{eqnarray}
E_D^{loop}=\frac{eg^2 A D(s)}{(2\pi)^4}\left[\int\frac{4((r+l)e)(l(q-p))d^4l}
{(l^2-m_\pi^2)((r+l)^2-m_\pi^2)((r+l-k)^2-m_\pi^2)}\right.
\nonumber \\
+\int\frac{4(le)(l(q-p))d^4l}
{(l^2-m_\pi^2)((l+k)^2-m_\pi^2)((r+l)^2-m_\pi^2)}\nonumber \\
\left.-2\int\frac{((q-p)e)d^4l}
{(l^2-m_\pi^2)((r+l)^2-m_\pi^2)} \right]=\frac{eg^2
A}{\pi^2}F(s)D(s)\frac{T_E}{rk}
=ebT_E\label{eq. F},
\end{eqnarray}
with
\begin{eqnarray}
F(s)=\frac{1}{2}+\frac{s}{2rk}\left[\beta\mbox{Arth}\left(\frac{1}{\beta}\right)-\beta_0
\mbox{Arth}\left(\frac{1}{\beta_0}\right)\right] \nonumber\\
-\frac{m_\pi^2}{rk}\left(\mbox{Arth}^2\left(\frac{1}{\beta}
\right)-\mbox{Arth}^2\left(\frac{1}{\beta_0}\right)\right)
\nonumber \\
+\frac{i\pi}{2rk}
\left(\frac{s}{2}\left(\beta-\beta_0\right)
-2m_\pi^2\left(\mbox{Arth}\left(\frac{1}{\beta}\right)
-\mbox{Arth}\left(\frac{1}{\beta_0}\right)\right)\right)
, \label{eq.Fs}
\end{eqnarray}
where $s=(r-k)^2$, $\beta=\sqrt{1-4m_\pi^2/s}$,
$\beta_0=\sqrt{1-4m_\pi^2/m_K^2}$, $g$ is the
interaction constant of $\rho\pi\pi$. We take the amplitude of the
$\rho\rightarrow\pi\pi$
decay as the interaction constant $g$ and use the experimental result for it.

We note that since $F(s)$ is complex, the phase of the loop
contribution is not equal to the pion $P$-wave scattering phase.
The photon energy dependence of the $b$ phase ($arg(b)=\delta_b$) is shown
in Figure 3.

Heavy particles in the loop can also contribute to the electric
direct emission amplitude, though they do not produce any
additional phase. The possible intermediate states are
$\pi K$, $K\eta$ and $KK$. However, the $KK$ loop vanishes in the limit
$m_{K^0}=m_{K^+}$.  The contribution of these loops
can be calculated under the assumption that the
interaction constants are the same as in the pion loop case
($g_{K\pi\pi}=g_{K\pi K}=g_{KK\eta}=g_{KKK}$ and
the same for the interaction with $\rho$).

\section{The spectrum departure from pure}
\section*{bremsstrahlung}

Now we can use the results on the $K_{S}\rightarrow\pi^{+}\pi^{-}\gamma$
decay probability, obtained in the previous Section, to
estimate the departure of the photon spectrum from the pure bremsstrahlung.
We neglect the $CP$-violating magnetic direct
emission amplitude. The calculations are carried out in the $K_S$-rest
frame ($r=(m_K,0,0,0)$, $k=(\omega,\omega,0,0)$), where
$$
s=m_K^2-2m_K \omega, ~~~ \beta=\sqrt{1-\frac{4m_\pi^2}{m_K^2-2m_K
\omega}},
$$
$$
rk=m_K \omega.
$$
So for the double differential decay width with unpolarized photon
we obtain
\begin{eqnarray}
\frac{d\Gamma(K_S\rightarrow\pi^{+}\pi^{-}\gamma)}{d\omega
d\cos\theta}= \frac{2\alpha}{\pi}\frac{\beta^3}{\beta_0}
\left(1-\frac{2\omega}{m_K}\right)\sin^2\theta\Gamma(K_S\rightarrow\pi^{+}\pi^{-})
\nonumber \\
\times\left[\frac{1}{\omega(1-\beta^2\cos^2\theta)^2}
+\frac{m_K^4|a+b|^2\omega^3}{16|A|^2}
+\frac{Re(be^{-i\delta_0^0})\omega m_K^2}{2|A|(1-\beta^2\cos^2\theta)}
\right].
\label{eq.G}
\end{eqnarray}
Here $\theta$ is an angle between photon and
$\pi^{+}$ in the dipion rest frame. Integrating Eq. (\ref{eq.G})
over $\cos\theta$ between the limits $-1\leq \cos\theta\leq 1$
we obtain the following result for the differential decay width:
\begin{eqnarray}
\frac{d\Gamma(K_S\rightarrow\pi^{+}\pi^{-}\gamma)}{d\omega}=
\frac{2\alpha}{\pi}\frac{\beta^3}{\beta_0}\left(1-\frac{2\omega}{m_K}\right)
\Gamma(K_S\rightarrow\pi^{+}\pi^{-})\nonumber \\
\times
\left\{\frac{1}{\omega}\left[\frac{1+\beta^2}{2\beta^3}\mbox{ln}\frac{1+\beta}{1-\beta}
-\frac{1}{\beta^2}\right]+\frac{m_K^4\omega^3}{12}\frac{|b|^2}{|A|^2}\right.\nonumber
\\
\left.+\frac{\mbox{Re}((a+b)e^{-i\delta_0^0})\omega m_K^2}{2|A|}
\left[\frac{2}{\beta^2}-\frac{1-\beta^2}{\beta^3}\mbox{ln}\frac{1+\beta}{1-\beta}\right]\right\}.
\label{eq. spect}
\end{eqnarray}
The second and the third terms in the brackets govern the departure
of photon spectrum from the pure bremsstrahlung. We characterize
the departure of spectrum by the ratio
\begin{equation}
R=\frac{\left.\frac{d\Gamma}{d\omega}\right|_{interf}}{\left.\frac{d\Gamma}{d\omega}\right|_{IB}}.
\label{R}
\end{equation}
The ratio increases with the increase of $\omega$, and varies from
0.1\% at $\omega=50$Mev to 1\% at $\omega=160$Mev (see Fig. 4).

\section{Comparison with ChPT}

Let us compare the results on the interference branching ratio
in the $K_S$ decay obtained in Refs. \cite{AmbrI}, \cite{bib 3}
with the results presented above. In the framework of the ChPT the electric
direct emission amplitude is a sum of the loop contribution and
the counterterm contributions. The counterterms are needed in the ChPT to
reabsorb divergencies arising from loops at each order in momenta,
because the ChPT is a nonrenormalizable theory.

In general the loop contribution and the counterterm contributions are
separately scale-dependent. However, in this case the counterterm contributions
are scale independent and the loop contribution is finite, as it was shown in
Refs. \cite{AmbrI}, \cite{bib 3}. The similar result for the loop contribution
is obtained in the present paper (see equation (\ref{eq. F})).

Counterterm contributions don't depend on the photon energy contrary to the
loop contribution, as it is shown in Refs. \cite{AmbrI}, \cite{bib 3}
\begin{equation}
E_{ct}=\frac{eG_8 m_K^3}{4\pi^2 F_{\pi}}N_{E_1},
\end{equation}
where $G_8$ is the interaction constant of the $|\Delta S|=1$
non-leptonic weak Lagrangian in the framework of the ChPT. The
index $8$ is due to the fact that the Lagrangian transforms under
$SU(3)_L\times SU(3)_R$ as an $(8_L, 1_R)$ or $(27_L, 1_R)$. Only the octet
part of the Lagrangian was taken into account in Ref. \cite{bib
3}. The value of $G_8$ was determined from the experimental data
on the $K_S\rightarrow\pi^{+}\pi^{-}$ decay probability:
$A(K_S\rightarrow\pi^{+}\pi^{-})=2 G_8 F_{\pi} (m_K^2-m_{\pi}^2)$,
$|G_8|=9\times 10^{-6}~\mbox{Gev}^{-2}$ . $F_\pi$ is the constant of the pion
leptonic decay: $F_\pi=93.3~\mbox{Mev}$, $N_{E_1}$ is a sum of the counterterm
constants and should be fixed from the experimental data.

The loop contribution according to Refs. \cite{AmbrI}, \cite{bib 3}
is equal to
\begin{equation}
E_{loop}=-\frac{eG_8 m_K(m_K^2-m_\pi^2)}{8\pi^2 F_\pi} \left(4h_{\pi\pi}
+h_{\pi K}+h_{K\eta}\right),
\end{equation}
where $h_{\pi\pi}$, $h_{\pi K}$, $h_{K\eta}$ denote the
contributions of corresponding loops. Regarding the $\pi\pi$ loop
it was found in Refs. \cite{AmbrI}, \cite{bib 3} that it dominates
the loop contribution to the electric direct emission amplitude
and is equal to
\begin{eqnarray}
E^{\pi\pi}_{loop}=-\frac{eG_8 m_K(m_K^2-m_\pi^2)}{8\pi^2 F_\pi
\omega^2} \left\{
s\left[\beta\ln\left(\frac{1+\beta}{\beta-1}\right)
-\beta_0\ln\left(\frac{1+\beta_0}{\beta_0-1}\right)\right]\right.
\nonumber \\
\left.+m_K\omega+m_\pi^2\left[\ln^2\left(\frac{1+\beta_0}{\beta_0-1}\right)
-\ln^2\left(\frac{1+\beta}{\beta-1}\right)\right]\right\}. \label{eq.loop}
\end{eqnarray}
We notice that the expression (\ref{eq.loop}) is proportional to
$(m_K^2-m_\pi^2)$, i. e. to the weak vertex $K\pi\pi$ with pions on-shell in
the framework of the ChPT. So the results obtained in the ChPT confirm the
assumption we made while considering diagrams of Fig. 2. We took
$A(K_S\rightarrow\pi^{+}\pi^{-})$ as the interaction constant and
considered pions on-shell.

In Refs. \cite{AmbrI}, \cite{bib 3} the "interference branching ratio"
was calculated for different values of $N_{E_1}=1.15 k_f$,
where $k_f=0,~\pm 0.5,~\pm 1$, because the ChPT can not fix the
values of the counterterm contributions, they can only be fixed
experimentally.

In order to compare the results of the present paper
on the "interference branching ratio" with the results obtained
in the framework of the ChPT we present the inner bremsstrahlung and
interference contributions to the branching ratio of the
$K_S\rightarrow\pi^{+}\pi^{-}\gamma$ decay, for different values of
the $\omega$ cut, along with the results of Refs.
\cite{AmbrI}, \cite{bib 3} in Table 1. As it can be seen from
Table 1 in the case of $k_f=0$ the "interference branching ratio"
obtained in the present paper and the one obtained in the
framework of the ChPT are in agreement for the photon energy cut
$\omega>20$ MeV. However, for the photon energy cuts $\omega>50$ MeV
and $\omega>100$ MeV the discrepancy appears. It is due to the
fact that the photon spectra of these results differ. It can be
clearly seen from Fig. 5, where the result of the present
paper (solid curve) along with the results of Refs. \cite{AmbrI},
\cite{bib 3} (dashed curves, $k_f=-0.5,~0,~+0.5$) for the photon
spectra of the interference contribution are shown.

If the counterterm contributions are switched on, then the arising
discrepancy is rather large for given in Refs. \cite{AmbrI},
\cite{bib 3} values of the counterterm contributions (e. g. for
$k_f=0.5$ the "interference branching ratio" obtained in
the present paper is 1.7 times larger than the one obtained in
Refs. \cite{AmbrI}, \cite{bib 3}; for $k_f=-0.5$ it is 3.3
times larger, see Table 1). It should be stressed that the
counterterm contributions don't depend on the photon energy and
exhibit the behavior different from the results of the present
paper (see Fig. 5). The obtained discrepancy is due to the fact
that in Refs. \cite{AmbrI}, \cite{bib 3} the $\rho$-meson
contribution shows up only in low-energy constants, while the
resonance contribution was not considered.. The phase of
the electric direct emission amplitude was taken to be $\delta_1^1(m_K)$,
the phase of $\pi\pi$ scattering in $P$-wave at energy $\sqrt{s}=_K$.
Instead we used the simple realistic model of $\pi\pi$ scattering via
$\rho$-meson taking into account the energy dependence of the $\delta_1^1(s)$
phase.

Indeed, in the case of the $\pi\pi$ scattering in framework of the ChPT the 
$\rho$-meson shows up in low-energy constants and as a direct resonance 
\cite{bib gasser}, \cite{bijnens}. However, in papers on 
$K\rightarrow\pi\pi\gamma$ decays based on the ChPT approach only low-energy
constants were accounted for. Such approach did not take into account the
energy dependence of the $\delta_1^1(s)$ phase, in other words the behavior
of the $\rho$-meson propagator, because the contributions of low-energy
constants do not depend on the energy. Thus, some dynamical features are 
missing in the ChPT approach for the $K\rightarrow\pi\pi\gamma$ decays. As it 
can be seen from Fig. 5 the standard ChPT approach even with higher order 
counterterms taken into account cannot reproduce the contribution of the 
$\rho$-meson.

In order to show that the inclusion of the $\rho$-meson into the
analysis of the electric direct emission in the $K\rightarrow\pi\pi\gamma$
decays is important, we address the the problem of $\pi\pi$
scattering in the $P$-wave in the following section.

\section{Pion-pion scattering}

In order to confirm our assumption that the simple realistic model for
describing the $\pi\pi$ interaction in the $P$-wave is more
appropriate than the standard ChPT approach,
let us compare the results of different models for $\pi\pi$
scattering with experimental data. In framework of the ChPT the amplitude of
$\pi\pi$
scattering to one loop takes the form \cite{bib gasser}
\begin{equation}
A(s,t,u)=\frac{s-M^2}{F_\pi^2}+B(s,t,u)+C(s,t,u)+O(p^6),
\end{equation}
where
\begin{eqnarray}
B(s,t,u)=(6F_\pi^4)^{-1}\left\{3(s^2-M^4)K(s)\right. \nonumber \\
+[t(t-u)-2M^2t+4M^2u-2M^4]K(t)\nonumber \\
\left.+[u(u-t)-2M^2u+4M^2t-2M^4]K(u)\right\}, \\
C(s,t,u)=(96\pi^2F_\pi^4)^{-1}\left\{2\left(l_1-\frac{4}{3}\right)(s-2M^2)^2)\right.\nonumber
\\
\left.+\left(l_2-\frac{5}{6}\right)[s^2+(t-u)^2]-12M^2s+15M^4\right\},
\end{eqnarray}
and
$$
K(q^2)=\frac{1}{16\pi^2}\left(\sigma\ln\frac{\sigma-1}{\sigma+1}+2\right),
$$
$$
\sigma=\left(1-\frac{4M^2}{q^2}\right)^{1/2}.
$$
This representation involves four constants: $F_\pi$, defined
above, $M$, pion mass, $l_1=0.4\pm0.3$ and $l_2=1.2\pm0.4$, which are
extracted from $\pi\pi$ data \cite{bib pp} and $K_{e4}$ decay \cite{bib Ke4}.

The two-loop representation of the scattering amplitude yields the first three
terms in the chiral expansion of the partial waves \cite{bijnens}:
\begin{equation}
t_l^I(s)=t_l^I(s)_2+t_l^I(s)_4+t_l^I(s)_6+O(p^8).
\end{equation}
This representation involves 12 constants. The leading order contains $F_\pi$
and
$M$, the next-to-leading order -- $l_1,~l_2,~l_3,~l_4$. The contribution of
the last two constants were not included to the scattering amplitude of Ref.
\cite{bib gasser},
though they appear in the next-to-leading order. This fact is due to
the smallness of the contributions proportional to these constants.
The next-to-next-to-leading order generates six coupling constants
$r_1,~...,~r_6$.

The two different categories of these constants should be distinguished.
First, the terms that survive in the chiral limit ($l_1,~l_2,~r_5,~r_6$).
They can be determined from the experimental data, as it
was mentioned above for $l_1$ and $l_2$. The constants $r_5(M_\rho)=3.8\pm1.0$,
$r_6(M_\rho)=1.0\pm0.1$ were calculated using the experimental data on $\pi\pi$
scattering and Roy equations in Ref. \cite{Col}.
Second, symmetry breaking terms. The corresponding vertices are proportional
to a power of the quark mass and involve the constants
$l_3,~l_4,~r_1,~r_2~r_3,~r_4$,
which may be determined by using other than $\pi\pi$ scattering experimental
information.
The constant $l_4=4.4\pm 0.2$ can be fixed by using experimental data on the
pion scalar form factor.
The contributions of $l_3$ to the scattering amplitude are very small and can
be neglected, as it
was shown in Ref. \cite{Col}. For $r_1,~...,~r_4$ the theoretical estimates
were
used in Ref. \cite{Col}.

The $\rho$-meson contribution is taken into account in the framework of the
ChPT
by considering the pole diagrams of $\pi\pi$-scattering via virtual
$\rho$-meson. This procedure
gives rise to an additional term in the expression for the amplitude of
$\pi\pi$ scattering, as it was shown in Refs. \cite{bib gasser},
\cite{bernard}.

In order to compare the amplitude with the experimental data on $P$-wave
$\pi\pi$ scattering \cite{bib alekseeva} we should expand the combination with
definite isospin in the s-channel
\begin{equation}
T^1(s,t)=A(t,u,s)-A(u,s,t)
\end{equation}
into partial waves with different angular momenta
$$
T^1(s,t)=32\pi\Sigma_{l=0}^{\infty}(2l+1)P_l(cos\theta)t_l^1(s),
$$
where $\theta$ is the scattering angle in the center-of-mass system of the
initial pions.
The partial amplitude is
\begin{equation}
t_1^1(s)=\frac{1}{64\pi}\int^1_{-1}T^1 P_1(\cos\theta)d\cos\theta.
\end{equation}

In Fig. 6 we present the behavior of the phase $\delta_1^1$ calculated in the
framework of the ChPT in one loop approximation (with and without $\rho$)
and the simple realistic model along with experimental data \cite{bib
alekseeva}.
We also present the behavior of the phase $\delta_1^1$ calculated in the
framework of the ChPT with $\rho$ in two-loop approximation in Ref. \cite{Col}.
According to the standard approach used when considering
phases of $\pi\pi$ scattering we utilize the elastic unitarity to
determine $\delta_1^1$ in the framework of the ChPT with and without
$\rho$. Namely
$$
Im(t^1_1)=\frac{2q}{\sqrt{s}}\left(Re(t_1^1)\right)^2,
$$
where $q^2=s/4-m_\pi^2$. Then the phase takes the following form
\begin{equation}
\delta_1^1=\arctan\left(\frac{2q}{\sqrt{s}}Re(t^1_1)\right). \label{d11}
\end{equation}

As it is seen from Fig. 6 the result obtained in the framework of the ChPT
without $\rho$ is in strict disagreement with the experimental data. It is due 
to the fact that the $\rho$ contribution was not taken into account.
In fact the use of equation (\ref{d11}) leads to the imaginary part
of bubble diagrams, which we considered above when calculating the pion loop
contributions to the $\rho$ propagator. Due to this fact the behavior of the
phases calculated in the framework of the ChPT with $\rho$ and in the framework
of the simple realistic model should be similar, as it is shown in Fig. 6.
However, the use of the unitarity to determine $\delta_1^1$ in the
framework of the ChPT with $\rho$ in one loop as well as in two loop
approximation leads to the imaginary part of the one-loop diagrams,
which describe the "resonance" structure of the $\rho$-meson propagator.
Instead, we considered the sum of the iterated bubble diagrams to calculate
the pion loop contributions to the $\rho$ propagator, as it is shown in 
Section 3. This fact leads to the difference between the result for the phase
obtained in the framework of the ChPT with $\rho$ contribution
taken into account and the result of the present paper, as it can
be seen in Fig. 6, which presents a magnified part of Fig. 6.
The two-loop approximation gives the result for  $\delta_1^1$
that is in better agreement with the result of the present paper.Thus, the 
inclusion of the $\rho$ meson into the analysis of the electric direct 
emission in the $K\rightarrow\pi\pi\gamma$ decays is important.

The theoretical curves shown in Figs. 6, 7 are determined with some uncertainty
due to the fact that we use the values obtained from the experimental data for
the interaction constants and the masses. However, the experimental errors of 
the interaction constants are very small, i. e. $F_\pi=93.3\pm0.3$, $g=6.08\pm 
0.03$, resulting in the uncertainty near $0.5\%$. The experimental results for 
the amplitude and the phase of the $\pi\pi$ scattering have the uncertainty 
near $15\%$, therefore we do not show the errors of the theoretical values in 
Figs. 6, 7.

\section{Analysis in terms of Stokes vectors}

For further calculations we will need the magnetic direct emission coupling
$c$. It
can be estimated using the experimental data on the direct emission
contribution to the $K_L\rightarrow\pi^{+}\pi^{-}\gamma$  decay \cite{bib
ramberg}, \cite{bib kettell}. The corresponding double differential
decay width for unpolarized photon is
\begin{equation}
\frac{d\Gamma(K_L\rightarrow\pi^{+}\pi^{-}\gamma)}{d\omega
d\cos\theta}= \frac{2\alpha}{\pi}\frac{\beta^3}{\beta_0}
\left(1-\frac{2\omega}{m_K}\right)
\frac{\Gamma(K_L\rightarrow\pi^{+}\pi^{-})c^2m_K^4}{16|A|^2|\eta_{+-}|^2}\sin^2\theta.
\end{equation}
Identifying this expression with the direct emission rate given in \cite{bib
ramberg},
\cite{bib kettell} we obtain $|c|=0.76|A|$.

It is worth mentioning that there is no interference between the
amplitude terms with opposite $CP$-parity, if photon polarization
is not observed. However, the interference is nonzero when the polarization is
measured, as it was repeatedly emphasized in Refs. \cite{bib 2a},
\cite{SW}, \cite{bib dolgov}, \cite{bib sanda}. Therefore, any $CP$-violation
involving interference of electric and magnetic amplitudes is
encoded in the polarization state of the photon.

To determine the nature of this interference we write the
$K_{L,S}\rightarrow \pi^{+}\pi^{-}\gamma$ decay amplitude more generally
as
\begin{equation}
A(K_{S,L}\rightarrow \pi^{+}\pi^{-}\gamma)=ET_E+MT_M,
\label{eq.1}
\end{equation}
where for the $K_S$ decay $E$ and $M$ have the form
\begin{equation}
E=\frac{eAe^{i\delta_0^0}}{(pk)(qk)}+eb~~\mbox{and}~~
M=ie\eta_{+-}c,
\end{equation}

and in the case of the $K_L$ decay we have
\begin{equation}
E=\eta_{+-}\left[\frac{eAe^{i\delta_0^0}}{(pk)(qk)}+eb\right]~~\mbox{and}~~
M=iec.
\end{equation}
The photon polarization can be defined in the terms of the density matrix
\cite{bib Ber}
\begin{equation}
\rho=\left(
\begin{array}{c}
|E|^2~~~~E^{*}M\\
EM^{*}~~~~|M|^2
\end{array}
\right)=\frac{1}{2}(|E|^2+|M|^2)\left[1+\bf{S\tau}\right],
\end{equation}
where $\bf{\tau}=(\tau_1,\tau_2,\tau_3)$ denotes Pauli matrices,
$\bf{S}$ is the Stokes vector of the photon with components
\begin{eqnarray}
S_1=2Re(E^{*}M)/(|E|^2+|M^2|), \\
S_2=2Im(E^{*}M)/(|E|^2+|M^2|), \label{eq.Stokes}\\
S_3=(|E|^2-|M|^2)/(|E|^2+|M^2|).
\end{eqnarray}
The effects of $CP$-violation reside in components $S_1$ and
$S_2$, while the component $S_3$ measures the relative strength of the
amplitude terms
with opposite $CP$-parity. The component $S_2$ is the net circular polarization
of the photon; it is proportional to the difference of $|E-iM|^2$ and
$|E+iM|^2$, which
are the probabilities for left-handed and right-handed polarization.
The $S_1$ component appears as a coefficient of an interference term in case of
linear polarization. If one chooses the polarization angle $\phi$ as
the angle between $\bf{e}$, the polarization vector, and the unit vector
$\bf{n}_\pi$ normal
to the decay plane (${\bf k}=(0,0,\omega)$, ${\bf n}_\pi=(1,0,0)$,
${\bf p}=(0,p\sin\theta,p\cos\theta)$), then the decay amplitude will be
proportional to the following expression
\begin{equation}
|A(K_{S,L}\rightarrow \pi^{+}\pi^{-}\gamma)|^2\sim
1-(S_3\cos 2\phi +S_1\sin 2\phi). \label{eq. lin}
\end{equation}

It is obvious from (\ref{eq. lin}) that the measurement of linear
polarization in principle allows to extract the terms with
opposite $CP$-parity ($E$ and $M$).
$$
|A(K_{S,L}\rightarrow \pi^{+}\pi^{-}\gamma)|^2\sim |E|^2,~~ \phi=\pi/2+\pi
n,
$$
$$
|A(K_{S,L}\rightarrow \pi^{+}\pi^{-}\gamma)|^2\sim |M|^2,~~ \phi=\pi n,
$$
where $n$ is integer.

In order to obtain a quantitative estimate of the $CP$-violation
effects we study the photon energy dependence of the Stokes vector
components. In figures 8 and 9 we show the photon energy dependence
of the $S_1$ and $S_2$ components in the
$K_{L,S}\rightarrow\pi^{+}\pi^{-}\gamma$
decay respectively. In figure 10 we demonstrate the $S_3$  component photon
energy
dependence in $K_{L,S}\rightarrow\pi^{+}\pi^{-}\gamma$ decays to
obtain the estimate of the relative strength of the $CP$-violation
effects in the decays under consideration. We see that the obtained results on
the
$K_L\rightarrow\pi^{+}\pi^{-}\gamma$ decay coincide with the
results of Refs. \cite{bib Sehgal1}, \cite{sehgal2}, \cite{sehgal3}. Taking
into account the
physical meaning of the $S_3$ we conclude that the $CP$-violation
effects in $K_S\rightarrow\pi^{+}\pi^{-}\gamma$ decay are
substantially
small. The reason is that the bremsstrahlung contribution shades
the magnetic direct emission contribution even for the high photon
energies.

It was suggested in Refs. \cite{bib sehgal}, \cite{heiliger}, \cite{elwood} to
use in place
of $\bf{e}$ the vector $\bf{n}_l$
normal to the $e^{+}e^{-}$ plane in the decay $K_L\rightarrow
\pi^{+}\pi^{-}e^{+}e^{-}$. This can be achieved by replacing
$e_\mu$ in the radiative amplitude (\ref{eq.2}, \ref{eq.3}) by
$e/k^2\overline{u}(k_{-})\gamma_\mu v(k_{+})$.
This motivates the study of the
distribution $d\Gamma/d\phi$ in the decays $K_{S,L}\rightarrow
\pi^{+}\pi^{-}e^{+}e^{-}$, where $\phi$ is an angle between
$\pi^{+}\pi^{-}$ and $e^{+}e^{-}$ planes.

The distribution $d\Gamma/d\phi$ can be written in general form
\begin{equation}
\frac{d\Gamma}{d\phi}=\Gamma_1 \cos^2\phi+\Gamma_2
\sin^2\phi+\Gamma_3 \sin\phi\cos\phi.
\end{equation}
The last term changes sign under the transformation $\phi\rightarrow\pi-\phi$
and
produces an asymmetry $A^{L,S}_{\pi\pi,~ee}$
in the distribution of the angle $\phi$
between the vectors normal to the $\pi^{+}\pi^{-}$ and $e^{+}e^{-}$
planes. The asymmetry is defined by the following expression
\begin{equation}
A^{L,S}_{\pi\pi,~ee}=\frac{\left(
\int^{\pi/2}_0-\int^{\pi}_{\pi/2}+\int^{3\pi/2}_{\pi}-\int^{2\pi}_{3\pi/2}\right)
\frac{d\Gamma}{d\phi}d\phi}
{\left(
\int^{\pi/2}_0+\int^{\pi}_{\pi/2}+\int^{3\pi/2}_{\pi}+\int^{2\pi}_{3\pi/2}
\right)\frac{d\Gamma}{d\phi}d\phi}.
\end{equation}

In the case of the $K_L$ decay the contributions of the amplitude
terms with different $CP$-parity to the decay probability are of
comparable magnitude. This fact should result in the significant
value of the asymmetry. This was demonstrated in Refs. \cite{bib
Sehgal1}, \cite{sehgal2}, \cite{sehgal3}, \cite{bib sehgal},
\cite{heiliger}, \cite{elwood}, \cite{Savage1}, \cite{Savage2}. The
$K_L\rightarrow\pi^{+}\pi^{-} e^{+}e^{-}$ decay probability and the
$CP$-violating asymmetry from the $\pi^{+}\pi^{-}$ and $e^{+}e^{-}$
planes correlation calculated in Refs. \cite{bib Sehgal1}, \cite{sehgal2},
\cite{sehgal3}, \cite{bib sehgal}, \cite{heiliger},
\cite{elwood}, \cite{Savage1}, \cite{Savage2}
\begin{equation}
Br(K_L\rightarrow\pi^{+}\pi^{-} e^{+}e^{-})=3.1\times10^{-7},
~~~|A^L_{\pi\pi,~ee}|=14\%,
\end{equation}
are in accordance with the recent experimental data published by KTeV
\cite{bib ktev}
($Br^{exp}(K_L\rightarrow\pi^{+}\pi^{-} e^{+}e^{-})=(3.32\pm0.14\pm0.28)
\times 10^{-7}$, $|A^L_{\pi\pi,~ee}|^{exp}=(13.6\pm2.5\pm1.2)\%$).

The $CP$-violating asymmetry $A^{L,S}_{\pi\pi,~ee}$ arises from the
interference of the amplitude terms with different $CP$-parity.
$A^{L,S}_{\pi\pi,~ee}$ is proportional to the expression
\begin{equation}
\int d\cos\theta_\pi ds dk^2 \sin^2 \theta_\pi \beta^3 X^2
\left(\frac{s}{k^2}\right)Re[ME^{*}],
\end{equation}
where
$$
X=\left[\left(\frac{m_K^2-s-k^2}{2}\right)-sk^2\right]^{1/2},
$$
$\theta_\pi$ is the angle between $\pi^{+}$ three-momentum and the
$K_L$ three-momentum in the $\pi^{+}\pi^{-}$ rest frame, $s=(p+q)^2$,
$k^2=(k_{+}+k_{-})^2$. The resulting $CP$-violating asymmetry in the
$K_L\rightarrow\pi^{+}\pi^{-} e^{+}e^{-}$ decay is
\begin{equation}
|A^L_{\pi\pi,~ee}|=(13.4\pm 0.9)\%.
\end{equation}
The value of the asymmetry is determined with some uncertainty due to the fact
that
we use the values obtained from the experimental data for interaction
constants, masses and
phase $\delta_0^0$. The central values of the asymmetry obtained in the
present paper and in
Refs. \cite{bib Sehgal1}, \cite{sehgal2}, \cite{sehgal3}, \cite{bib sehgal},
\cite{heiliger}, \cite{elwood}, \cite{Savage1}, \cite{Savage2} coincide
within the accuracy of the calculation.

According to the results of the present paper
the $CP$-violating asymmetry in the $K_S\rightarrow\pi^{+}\pi^{-} e^{+}e^{-}$
decay is
\begin{equation}
|A^S_{\pi\pi,~ee}|=(5.1\pm 0.4)\times 10^{-5}.
\end{equation}
This value of the asymmetry could be expected qualitatively from
the analysis of the $K_S\rightarrow\pi^{+}\pi^{-}\gamma$ decay
amplitude, where the $CP$-violating magnetic direct emission
contribution is substantially small compared to the
$CP$-conserving part of the amplitude.

\section{Conclusions}

In the case  of the radiative $K$-meson decays we calculated
the phases of amplitude terms using a simple realistic model
of pion-pion interaction \cite{bib Lee}. Also we
calculated the pion loop contribution ($E_D^{loop}$) to the
electric direct emission amplitude. The interference of the $E_D^{loop}$ with
the bremsstrahlung contribution is the main source of the departure of
the photon spectrum from pure bremsstrahlung. To detect this
effect the photon spectrum should be measured with the
accuracy better than 1\% for photon energies near 160 Mev and
better than 0.1\% for photon energies near 50 Mev.

We compared our results on the interference contribution to the
$K_S\rightarrow\pi^+\pi^-\gamma$ decay probability with
those of Refs. \cite{AmbrI}, \cite{bib 3} and found that 
the "interference branching ratio" differs from the the one obtained in Refs. 
\cite{AmbrI}, \cite{bib 3} (see Table 1). If the counterterm contributions are 
set to zero ($k_f=0$), the  result of the present paper and the result 
obtained in the framework of the ChPT are in agreement for the photon energy 
cut $\omega>20$ MeV. However, for the photon energy cuts  $\omega>50$ MeV and 
$\omega>100$ MeV the discrepancy appears. It is due to the fact that the 
photon spectra of these results differ. It can be clearly seen from Fig. 5, 
where the result of the present paper (solid curve) along with the results of 
Refs. [5], [15] (dashed curves, $k_f=-0.5,~0,+0.5$) for the photon spectra of 
the interference contribution are shown.

If the counterterm contributions are switched on, then the 
arising discrepancy is rather large for given in Refs. [5], [15] values of the 
counterterm contributions (e.g. for $k_f=0.5$ the result of the present paper 
is 1.7 times smaller that the one of Refs. [5], [15]; for $k_f=-0.5$ it is 3.3 
times larger, see Table 1). It should be stressed that the counterterm 
contributions do not depend on the photon energy and exhibit the behavior 
different from the results of the present paper (see Fig. 5). This discrepancy 
arises from different models of $\pi\pi$ interaction and the fact that we took 
into account the energy dependence of the phases, while in the ChPT approach 
the phases were taken at the energy $\sqrt{s}=m_K$. Thus it is clear that the 
standard ChPT approach for the $K\rightarrow\pi\pi\gamma$ decays even with 
higher order counterterms taken into account cannot reproduce the contribution 
of the $\rho$ meson. 

Actually, in the case of $\pi\pi$ scattering in the framework of the ChPT the 
$\rho$-meson shows up in two ways: as low-energy constants and as a direct 
resonance. However, in papers on the $K\rightarrow\pi\pi\gamma$ decays only 
low-energy constants are accounted for. Such approach doesn't take into 
account the energy dependence of the $\delta^1_1(s)$ phase, i. e. the behavior 
of the $\rho$ propagator, because the contribution of low-energy constants 
doesn't depend on the photon energy. Thus, some dynamical features are missing 
in the ChPT approach for the $K\rightarrow\pi\pi\gamma$ decays. 

In order to show that the inclusion of the $\rho$ meson into the analysis of 
the electric direct emission in the $K\rightarrow\pi\pi\gamma$ decays is 
important, we addressed the problem of the $\pi\pi$ scattering in $P$-wave.
We compared  with the experimental data the predictions for the phase of the 
$\pi\pi$ scattering in the $P$-wave obtained in the framework of different 
models. As seen from Fig. 6 the simple realistic model and the ChPT with $\rho$
are in accordance with the experimental data. The ChPT without $\rho$ shows 
strong disagreement with data. In Fig. 7 we present the behavior of the phase 
at low energies. The ChPT with $\rho$ and our simple realistic model 
predictions differ due to the fact that actually the one-loop approximation 
was used to describe the $\rho$-meson contribution to the phase obtained in 
the framework of the ChPT with $\rho$, while we summed the iterated bubble 
diagrams to obtain the phase. Thus we can conclude from the analysis of the 
$\pi\pi$ scattering in the $P$-wave that the inclusion of the $\rho$ meson, 
done in the present paper, in the $K\rightarrow\pi\pi\gamma$ decays is 
important.

Regarding the dependence of the $K_S$ decay probability on
photon polarization we found that the measurement of the linear polarization
allowed in principle extraction of terms with opposite $CP$-parity.

We also studied the $K_{S,L}\rightarrow\pi^{+}\pi^{-} e^{+}e^{-}$
decays. The $CP$-violating asymmetry in the case of the $K_L$ decay was
found to be $(13.4\pm 0.9)\%$. The central values of the asymmetry obtained
in the present paper and in Refs. \cite{bib Sehgal1}, \cite{sehgal2},
\cite{sehgal3}, \cite{bib sehgal}, \cite{heiliger}, \cite{elwood},
\cite{Savage1}, \cite{Savage2} coincide within the accuracy of the calculation.
We found that the $CP$-violating asymmetry in the case of the $K_S$
decay is substantially smaller than in the $K_L$ case being equal to $(5.1\pm
0.4)\times 10^{-5}$, as it could be expected from the analysis of the
$K_S\rightarrow\pi^{+}\pi^{-}\gamma$ decay.

\section*{Acknowledgments}

The author appreciates L. B. Okun's scientific supervision and formulation of
this
problem. The author would also like to thank R. B. Nevzorov for fruitful
discussions and G. D'Ambrosio, M. I. Vysotsky and E. P. Shabalin for valuable
remarks. This work was
supported by the RFBR (grant N 96-02-18010).

\newpage

\section*{Figure captions}

\noindent Fig. 1. The interaction of pions in the case of the internal
bremsstrahlung.

\bigskip

\noindent Fig. 2. The emission of the photon from the loops of virtual
particles ($D$).

\bigskip

\noindent Fig. 3. The photon energy dependence of the pion loop phase
$\delta_b$ ($b=|b|e^{i\delta_b}$), in degrees.

\bigskip

\noindent Fig. 4. The photon energy dependence of the ratio $R$ defined in
(\ref{R})

\bigskip

\noindent Fig. 5. The photon energy dependence of the interference
contribution. The result of the present paper is represented by
solid curve. The results of ChPT -- by dashed curves. The upper
dashed curve corresponds to $k_f=-0.5$. The lower
dashed curve -- to $k_f=0.5$. The dashed curve in between
-- to $k_f=0$.

\bigskip

\noindent Fig. 6. The energy dependence of the $\pi\pi$ scattering
phase $\delta_1^1(s)$ in the $P$-wave, in degrees. The experimental data
are represented by stars. The ChPT
without $\rho$-meson contribution -- by diamonds.  The ChPT with the
$\rho$-meson
contribution taken into account, according to the equations (3.11 -- 3.14) of
Ref. \cite{bernard}
-- by filled triangles. The ChPT with $\rho$ in two-loop approximation,
according to the result of Ref. \cite{Col} -- by empty triangles.
The results of the present paper -- by boxes.

\bigskip

\noindent Fig. 7. The energy dependence of the $\pi\pi$ scattering
phase $\delta_1^1(s)$ in the $P$-wave at low energies, in degrees.
Note that this figure is a magnified part of the previous figure.
The notations are the same as in Fig. 5.

\bigskip

\noindent Fig. 8. Stokes parameters $S_1$ (upper curve) and $S_2$ (lower
curve) for the
$K_L\rightarrow\pi^{+}\pi^{-}\gamma$ decay.

\bigskip

\noindent Fig. 9. Stokes parameters $S_1$ (upper curve) and $S_2$ (lower
curve) for the
$K_S\rightarrow\pi^{+}\pi^{-}\gamma$ decay.

\bigskip

\noindent Fig. 10. Stokes parameter $S_3$ for the
$K_S\rightarrow\pi^{+}\pi^{-}\gamma$ decay (straight line $S_3=1$) and
for the $K_L\rightarrow\pi^{+}\pi^{-}\gamma$ decay (lower curve).

\newpage

\begin{tabular}{|l|l|l|l|l|}
\hline
cut~~in $\omega$& $\omega >20$\ \mbox{Mev} & $\omega >50$\ \mbox{Mev} &
$\omega >100$\
\mbox{Mev}
\\ \hline
$10^3$\mbox{B}(the present paper) & 4.81 & 1.78 & 0.44 \\ \hline
$10^3$\mbox{B}~~ \cite{bib 3}&
4.80 & 1.73 & 0.31 \\ \hline
$10^6$ \mbox{Interf} & -6.3 & -4.8 & -1.7 \\ \hline
$10^6$\mbox{Interf}($k_f=0$) ~~\cite{bib 3}& -6.2 & -5.0 & -2.0\\ \hline
$10^6$\mbox{Interf}($k_f=0.5$) ~~\cite{bib 3}& -10.5 & -8.3 & -3.3\\ \hline
$10^6$\mbox{Interf}($k_f=1.0$) ~~\cite{bib 3}& -14.8 & -11.7 & -4.7\\ \hline
$10^6$\mbox{Interf}($k_f=-0.5$) ~~\cite{bib 3}& -1.9 & -1.6 & -0.6\\ \hline
$10^6$\mbox{Interf}($k_f=-1.0$) ~~\cite{bib 3}& +2.4 & +1.8 & +0.7\\ \hline

\end{tabular}

\begin{center}
Table 1

Internal Bremsstrahlung and interference contributions to the
branching ratios of the $K_S\rightarrow\pi^{+}\pi^{-}\gamma$ decay
for different $\omega$ cuts along with the results of Ref.
\cite{bib 3}
\end{center}

\newpage

\includegraphics[width=12cm]{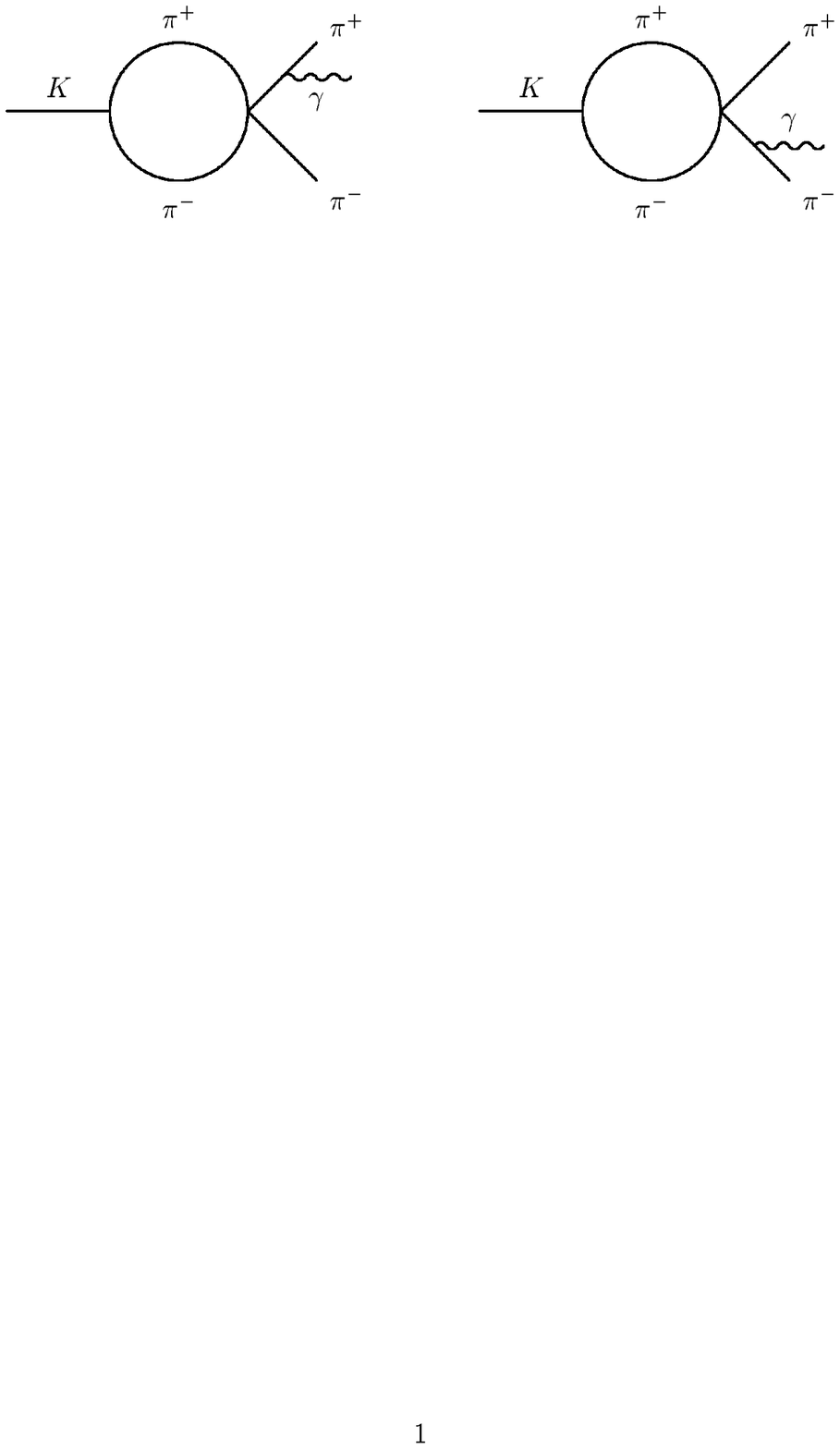}

\begin{center}
Fig. 1.
\end{center}

\includegraphics[width=12cm]{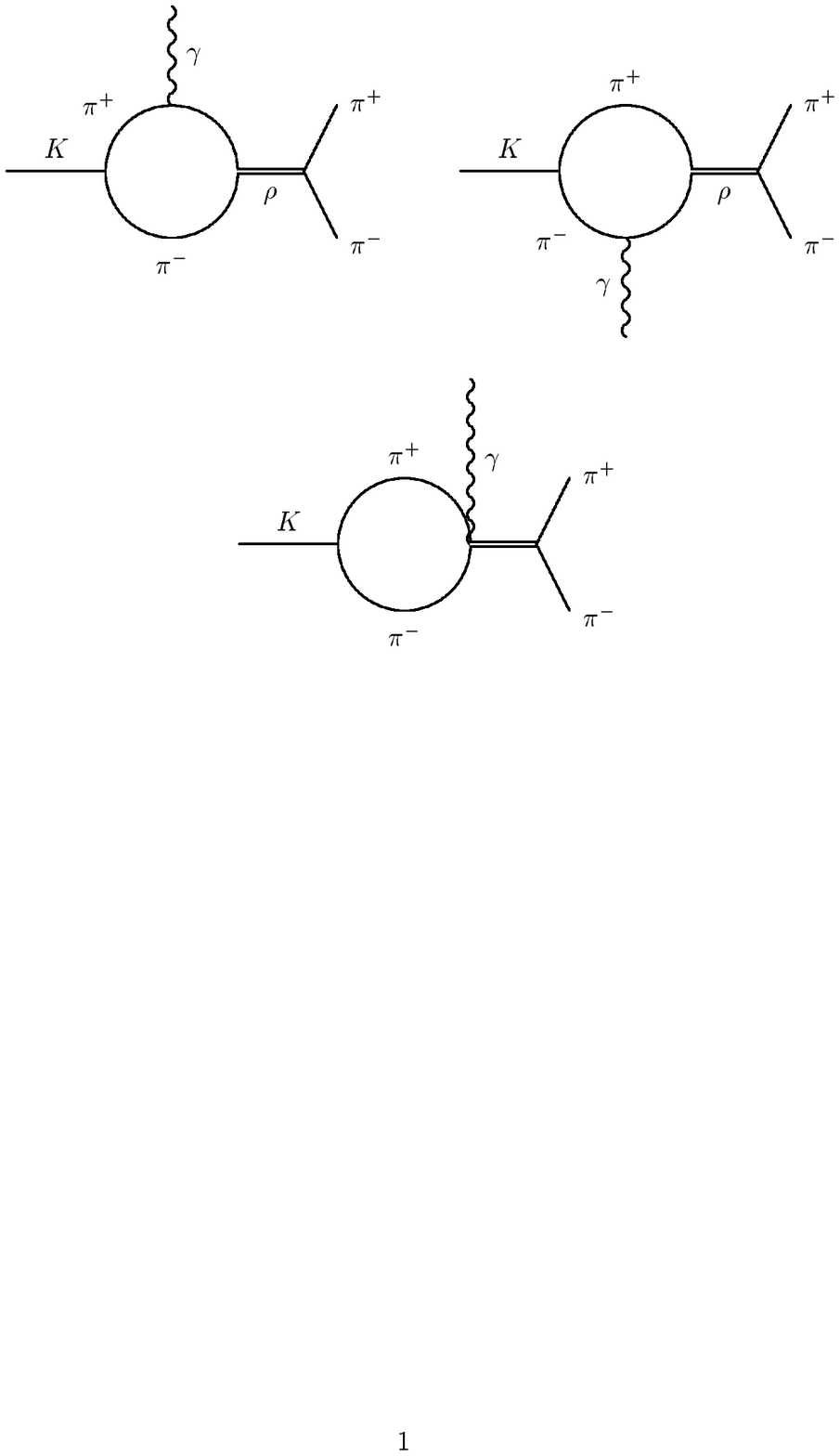}

\begin{center}
Fig. 2.
\end{center}

\newpage

\Large $\delta_{b}$

\includegraphics[width=10cm]{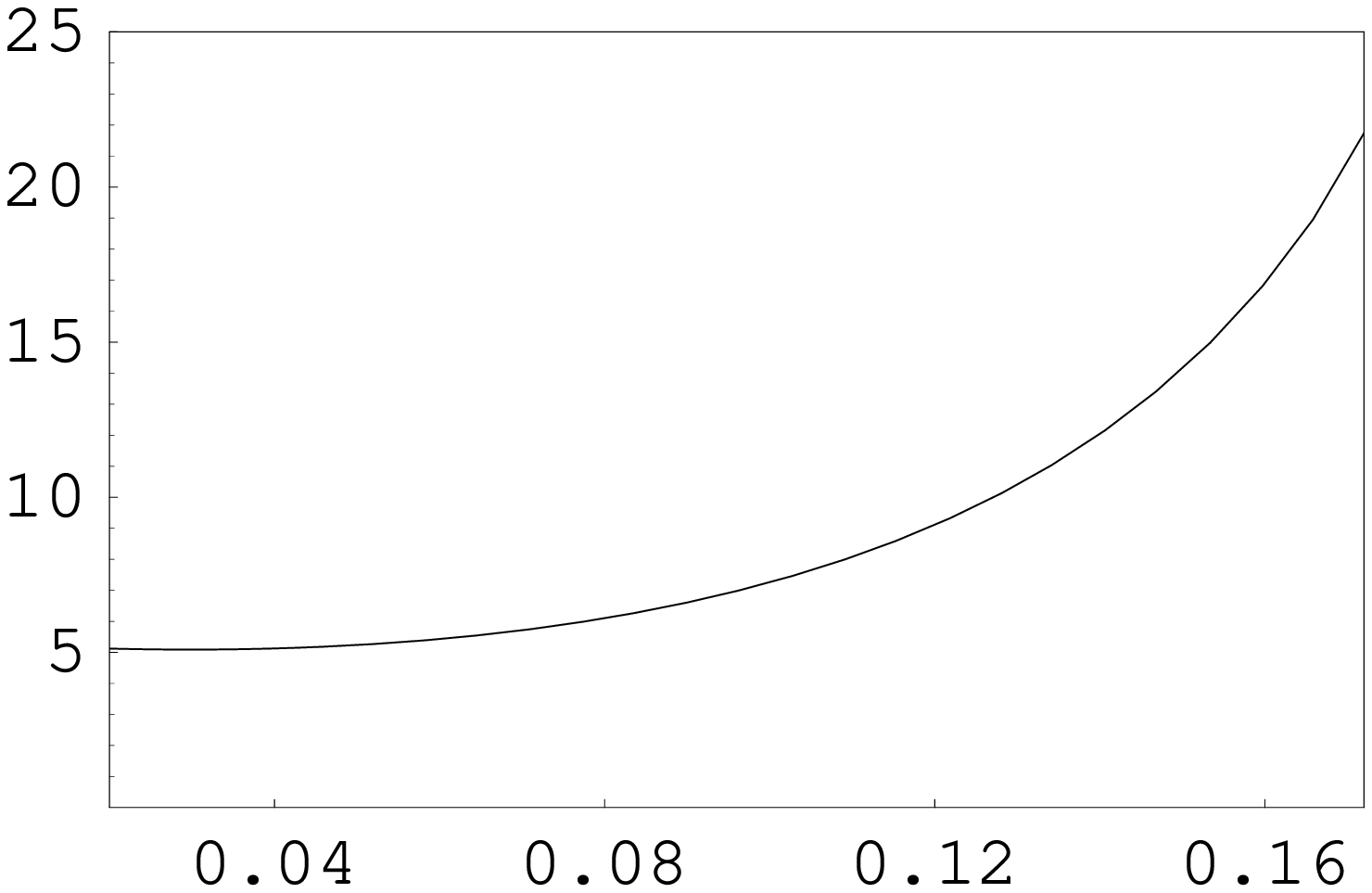}

\begin{flushright}
\Large $\omega$, GeV~~~~~~~~~~~~~
\end{flushright}

\begin{center}
Fig. 3.
\end{center}

\Large$R$

\includegraphics[width=10cm]{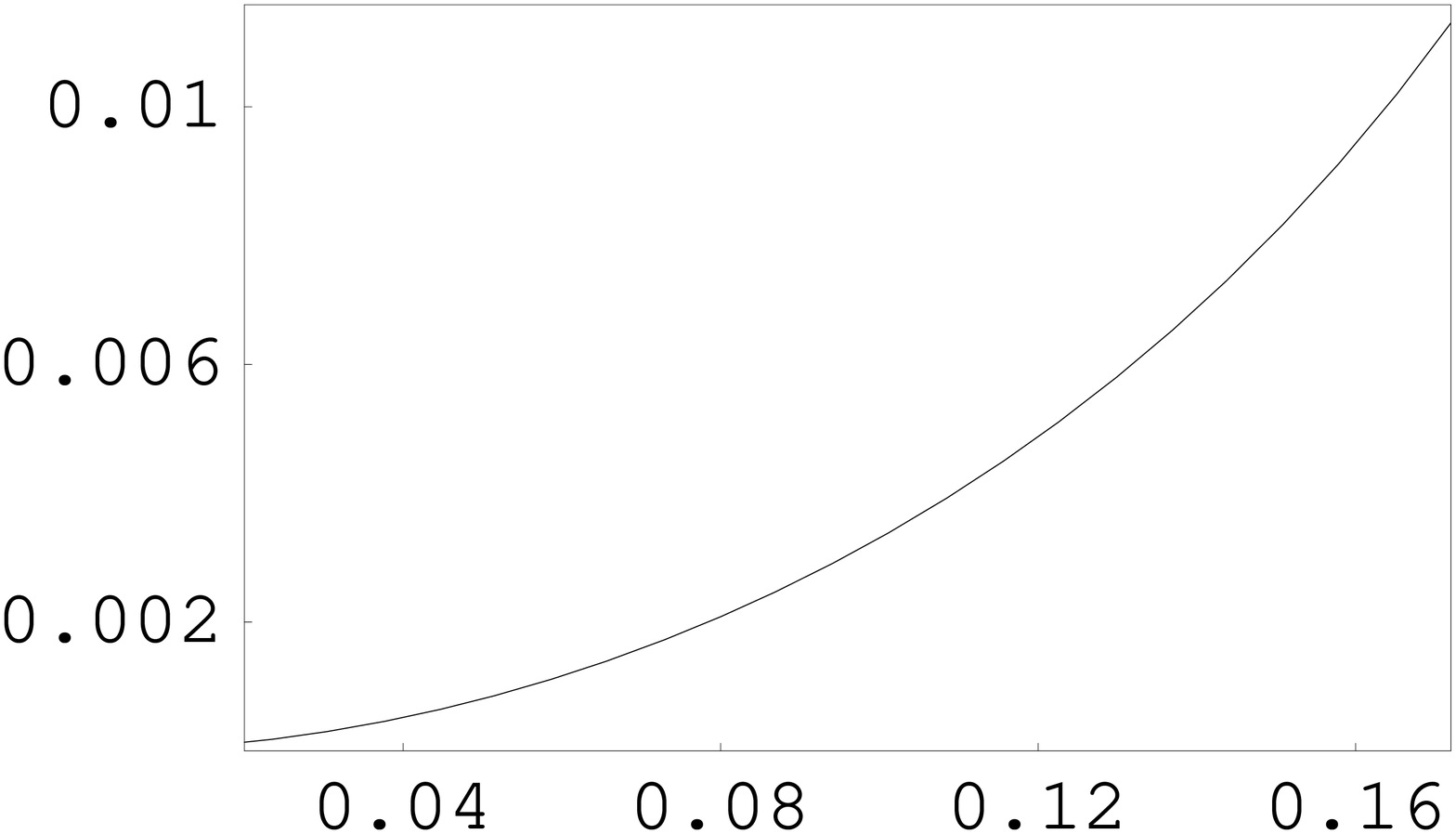}

\begin{flushright}
\Large $\omega$, GeV~~~~~~~~~~~
\end{flushright}

\begin{center}
Fig. 4.
\end{center}

\newpage

\Large $\left.\frac{d\Gamma}{d\omega}\right|_{interf}$

\includegraphics[width=10cm]{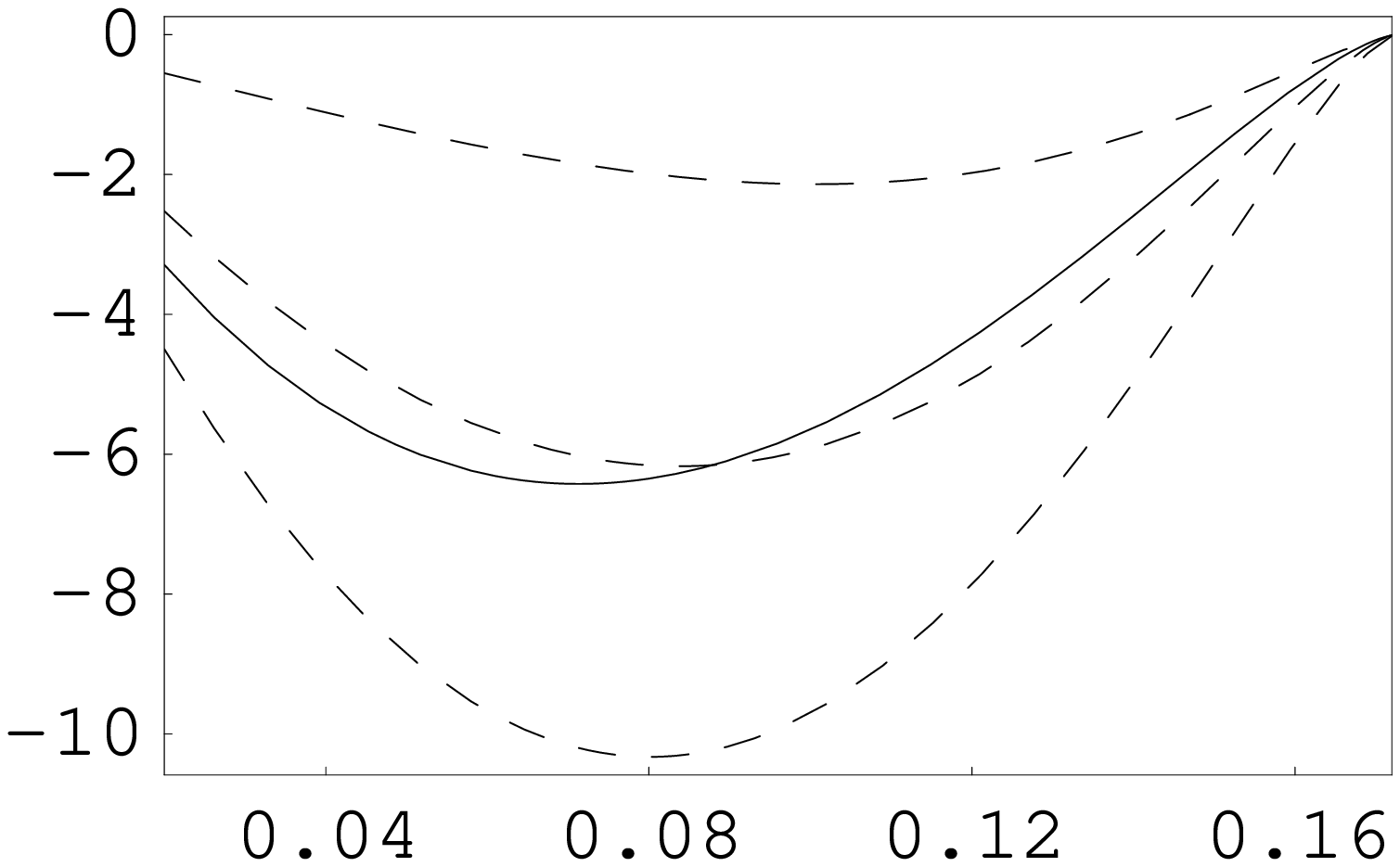}

\begin{flushright}
\Large $\omega$, GeV~~~~~~~~~~~~~~
\end{flushright}

\begin{center}
Fig. 5.
\end{center}

\Large $\delta^1_1$

\includegraphics[width=10cm]{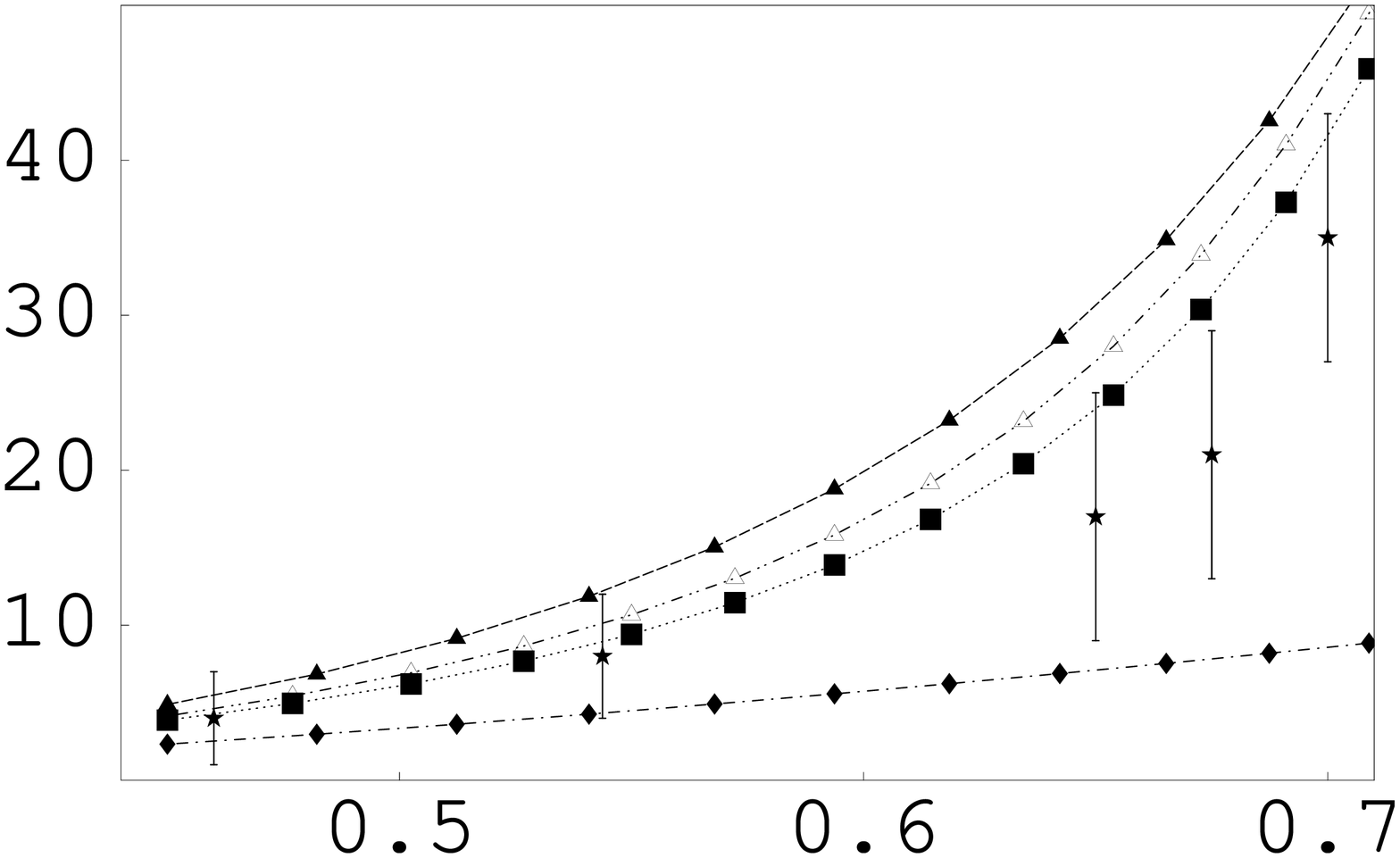}

\begin{flushright}
\Large s, GeV~~~~~~~~~~~~~~
\end{flushright}

\begin{center}
Fig. 6.
\end{center}

\Large $\delta_1^1$

\includegraphics[width=10cm]{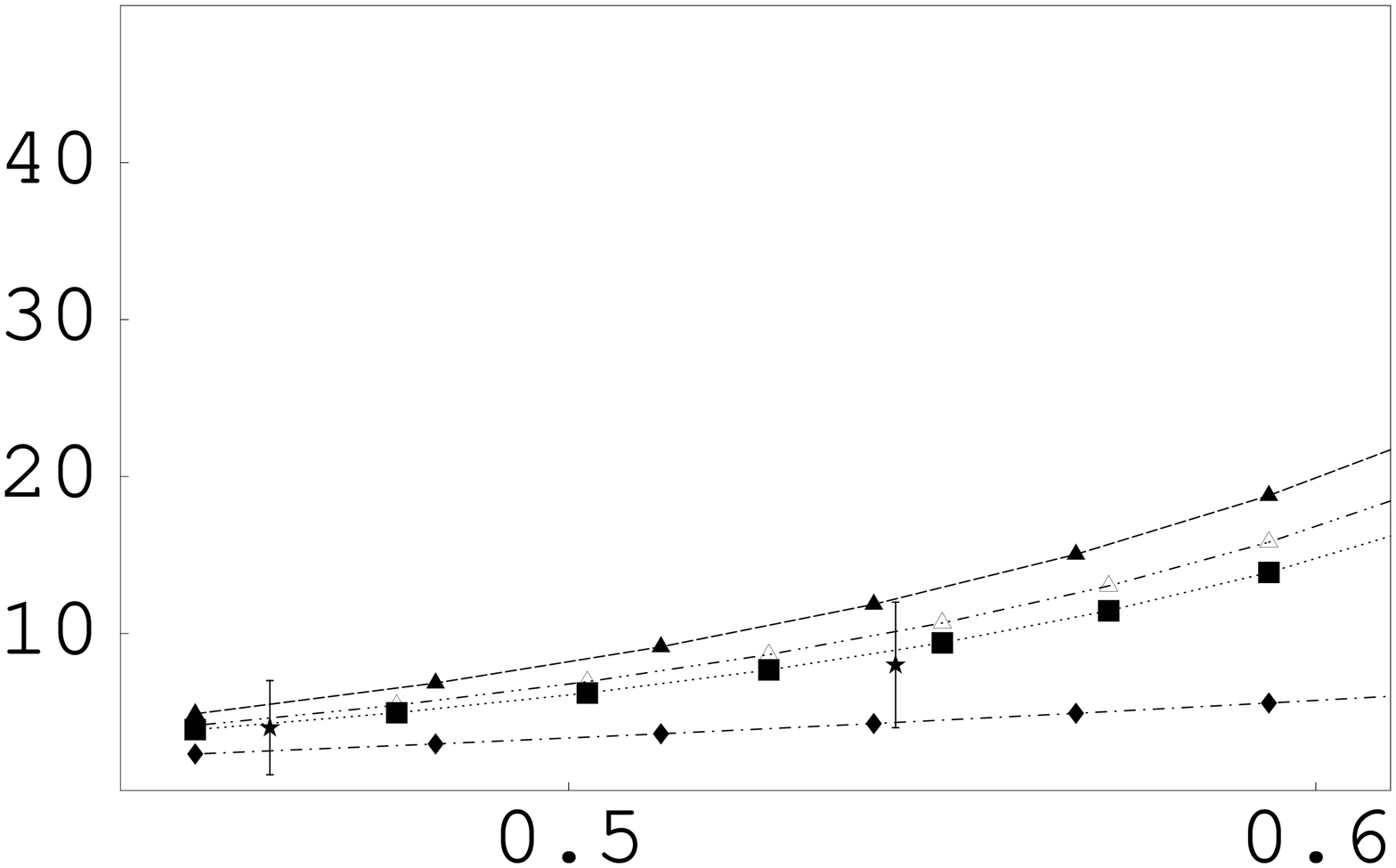}

\begin{flushright}
\Large s, GeV~~~~~~~~~~~~~~
\end{flushright}

\begin{center}
Fig. 7.
\end{center}

\Large $S_{1,2}$

\includegraphics[width=10cm]{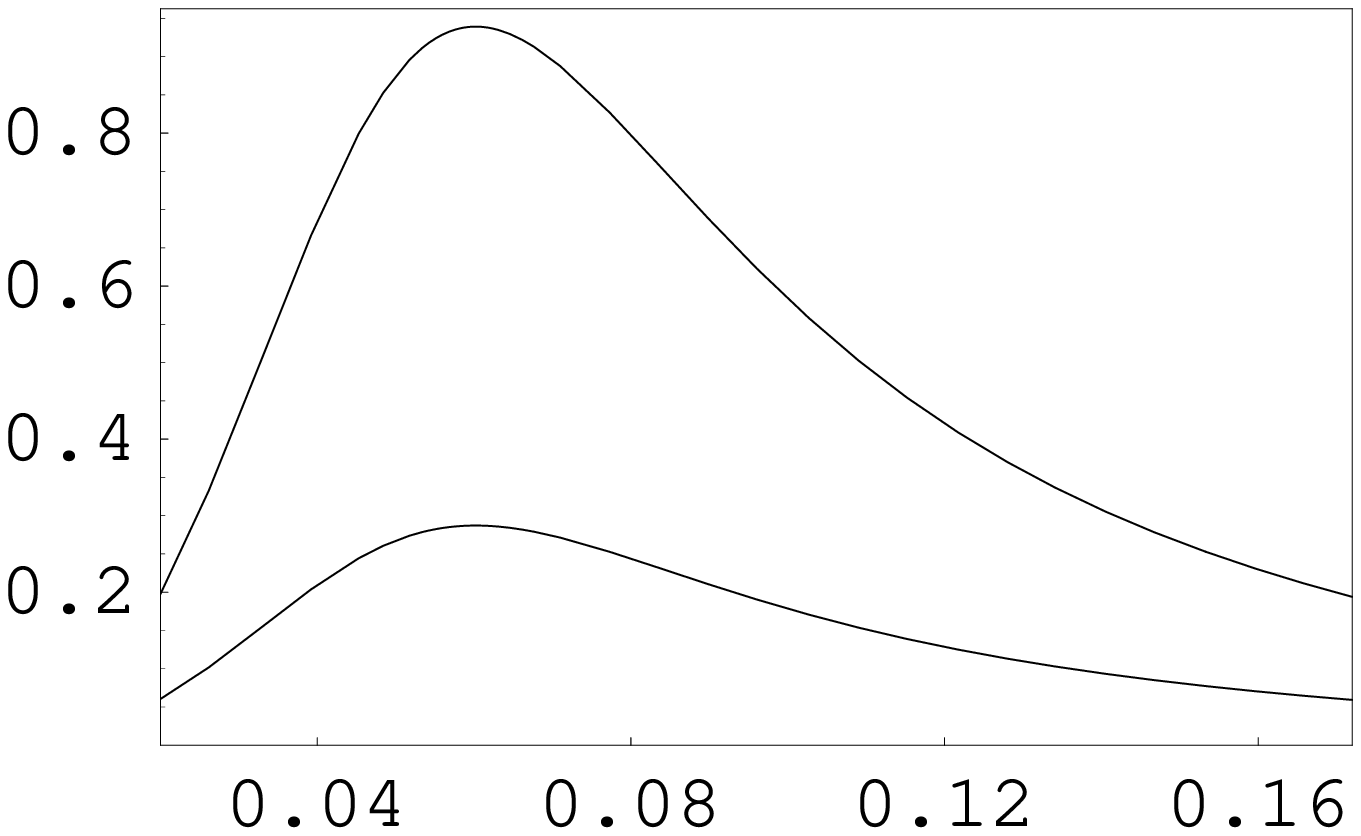}

\begin{flushright}
\Large $\omega$, GeV~~~~~~~~~~~~~
\end{flushright}

\begin{center}
Fig. 8.
\end{center}

\Large $S_{1,2}\times 10^{-4}$

\includegraphics[width=10cm]{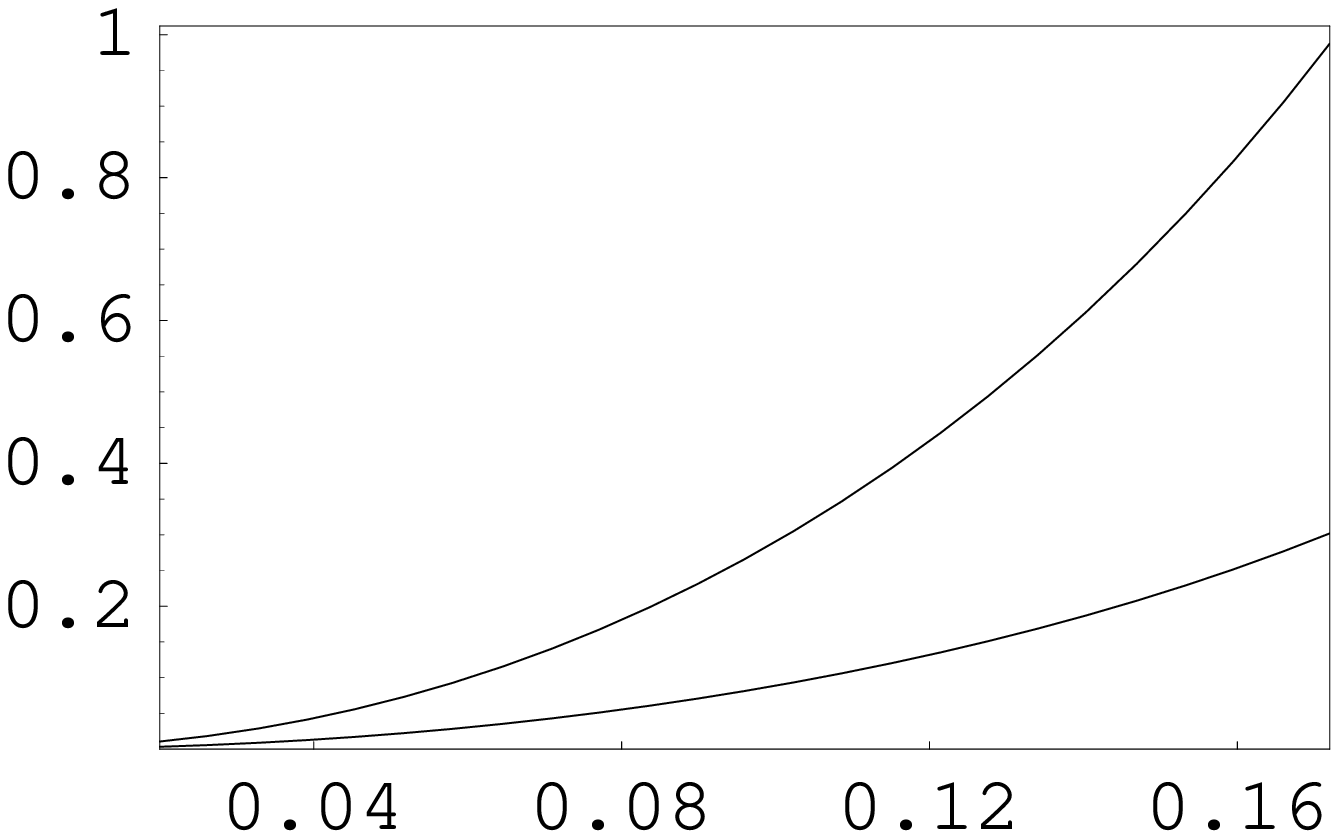}

\begin{flushright}
\Large $\omega$, GeV~~~~~~~~~~~
\end{flushright}

\begin{center}
Fig. 9.
\end{center}

$S_{3}$

\includegraphics[width=10cm]{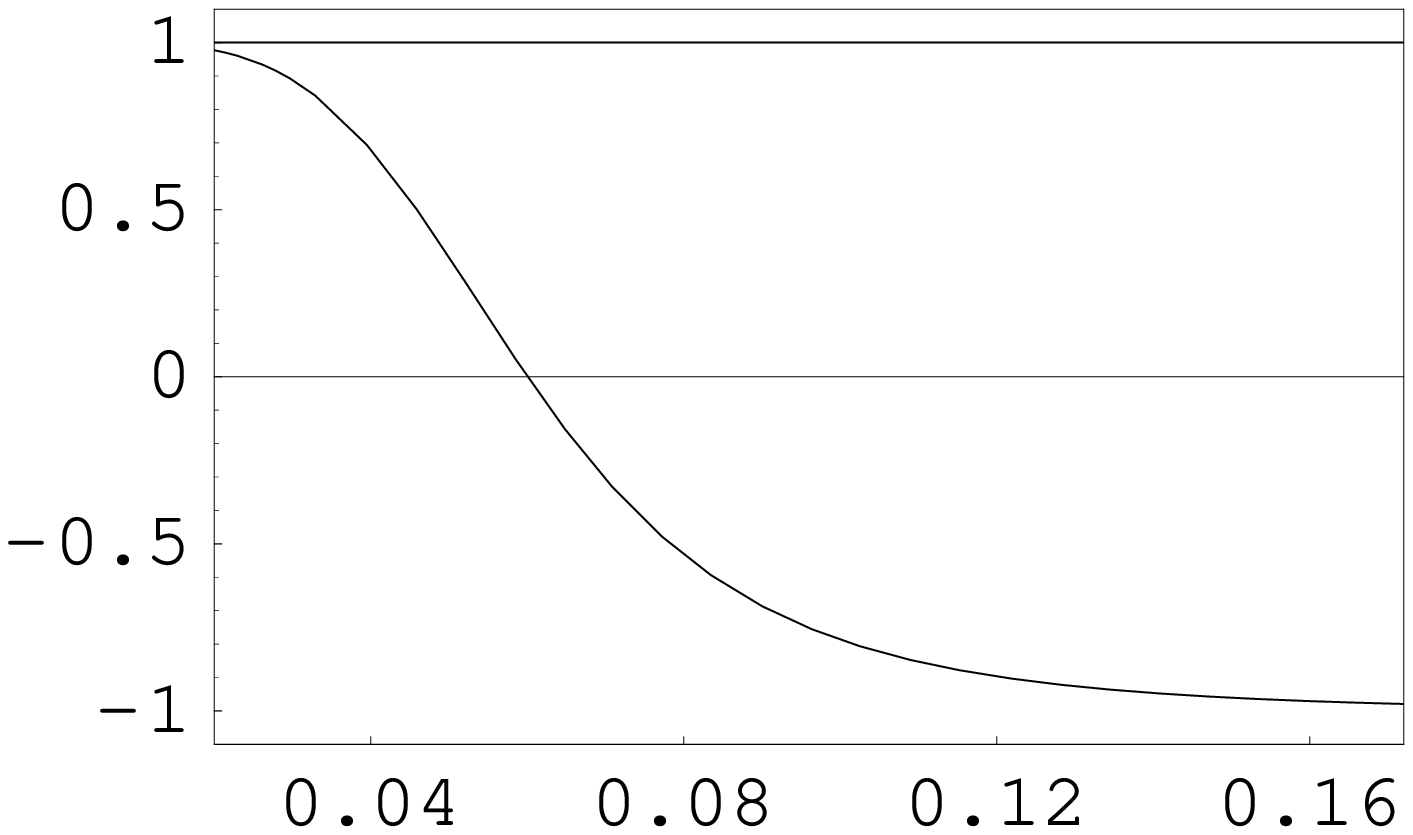}

\begin{flushright}
\Large $\omega$, GeV~~~~~~~~~~~
\end{flushright}

\begin{center}
Fig. 10.
\end{center}

\end{document}